\documentclass[twocolumn,amsmath,amssymb,showpacs,superscriptaddress]{revtex4}
\usepackage{graphicx}
\usepackage{dcolumn}
\usepackage{bm}
\usepackage{amsmath}
\usepackage{amsfonts}
\usepackage{amssymb}
\usepackage{here}
\makeatletter

\begin{document}
\title{Kinetic theory for dilute cohesive granular gases with a square well potential}
\author{Satoshi Takada\footnote{takada.satoshi.3s@kyoto-u.ac.jp. Present address: Department of Physics, Kyoto University, Kitashirakawa Oiwakecho, Sakyo-ku, Kyoto 606-8502, Japan}}
\affiliation{Yukawa Institute for Theoretical Physics, Kyoto University, Kitashirakawa Oiwakecho, Sakyo-ku, Kyoto 606-8502, Japan}
\author{Kuniyasu Saitoh\footnote{Present address: WPI Advanced Institute for Materials Research, Tohoku University, 2-1-1 Katahira, Aoba-ku, Sendai, 980-8577, Japan}}
\affiliation{Faculty of Engineering Technology, MESA+, University of Twente, 7500 AE Enschede, The Netherlands}
\author{Hisao Hayakawa}
\affiliation{Yukawa Institute for Theoretical Physics, Kyoto University, Kitashirakawa Oiwakecho, Sakyo-ku, Kyoto 606-8502, Japan}
\date{\today}

\begin{abstract}
We develop the kinetic theory of dilute cohesive granular gases in which the attractive part is described by a square well potential.
We derive the hydrodynamic equations from the kinetic theory with the microscopic expressions for the dissipation rate and the transport coefficients. 
We check the validity of our theory by performing the direct simulation Monte Carlo.
\end{abstract}
\maketitle

\section{introduction}
The hydrodynamic description of granular materials is useful to know the rheological properties of the granular flow.
Since granular materials are recognized to behave as unusual solids, liquids and gases,
granular materials have attracted much interest among physicists \cite{Jaeger1996}.
The most idealistic granular system is a dilute gas without any external forces such as gravity.
To analyze such a simple system is important to understand complex behavior of granular materials.
If the kinetic energy or the granular temperature of a granular gas homogeneously decreases because of inelastic collisions between grains, 
the time evolution of the temperature obeys Haff's law \cite{Haff1983}.
However, this homogeneous cooling state cannot be maintained as time goes on, 
because clusters of dense region appear \cite{Goldhirsch1993,Goldhirsch1993-2,McNamara1996}.
Such inhomogeneity of granular gases can be understood by granular hydrodynamics \cite{Brey1999,Savage1992,Garzo2006,Saitoh2011,Saitoh2013} in which the transport coefficients for the inelastic hard core system 
for the dilute case \cite{Lun1984, Jenkins1985, Noije1998, Brey1998, Huthmann2000} and the moderately dense case \cite{Resibois, Garzo1999} can be determined by the inelastic Boltzmann-Enskog equation \cite{Brey1998,Brey1999,Goldshtein1995,Sela1998}.
These theoretical results exhibit good agreements with the numerical simulations, at least, for nearly homogeneous moderately dense granular flows \cite{Mitarai2007,Chialvo2013}.
It should be noted that we often use the direct simulation Monte Carlo (DSMC) to evaluate the transport coefficients instead of using the molecular dynamics simulation, 
which was originally introduced by Bird \cite{Bird} to study rarefied gas \cite{Alexander1997, Garcia, Nanbu1980, Nanbu1983} and later has been extended to dilute inelastic gases \cite{Brey1999, Brey1999_2} and to dense inelastic gases \cite{Montanero2004, Montanero2005}.
This is because we should keep the system almost uniform.

The interaction between contacting granular particles usually consists of the repulsive force and the dissipative force proportional to the relative speed.
For fine powders and wet granular particles, however, cohesive force cannot be ignored.
The origins of such cohesive force are, respectively, van der Waals force for fine powders and capillary force for wet granular particles \cite{Rowlinson,Castellanos2005,Mitarai2006}.
Such cohesive forces can cause the liquid-gas phase transition \cite{Saitoh2015},
the variations of cluster formation of freely falling granular particles \cite{Ulrich2009-1,Ulrich2009-2,Ulrich2012,Royer2009,Waitukaitis2011, Weber2004}, and
the enhancement of the jamming transition \cite{Gu2014,Irani2014}.
Thus, the study of cohesive granular materials is important for both physics and industry to treat real granular materials.
In our previous paper, we have demonstrated the existence of various phases for fine powders in the presence of a plane shear, which cannot be observed in granular gases under the shear \cite{Takada2014}.
We have also developed the dynamic van der Waals model in describing such a system \cite{Saitoh2015} and obtain qualitatively consistent results with those in Ref.\ \cite{Takada2014}.
These results suggest that the ordinary kinetic theory for a hard core system cannot be applied to this system.
Needless to say, the kinetic theory is important to give us the microscopic basis of the macroscopic phenomenology such as Ref.\ \cite{Saitoh2015} and the simulation results such as Ref.\ \cite{Takada2014}.
In this paper, let us consider a granular gas whose interaction consists of the hard core for repulsive part and a square well potential for an attractive part.
There exist some studies on the kinetic theory of gas molecules having the square well potential 
\cite{Holleran1951, Longuet-Higgins1958, Davis1961, Davis1965, Altenberger1975, Polewczak2001}
in which, the collision processes are categorized into four processes: (i) hard core collisions, (ii) entering processes, (iii) leaving processes from the well, and (iv) trapping processes by the well \cite{Davis1961,Davis1965,Karkheck1985}.
Note that most of previous works study gases without dissipations in collisions except for some recent papers \cite{Muller2011, Murphy2015},
which do not discuss the transport coefficients.
It should also be noted that some papers developed the kinetic theory based on different models for cohesion \cite{Gidaspow1998, Kim2002}.

In this paper, we derive modified Haff's law and derive the transport coefficients for the dilute cohesive granular gases in freely cooling processes.
For this purpose, we extend the kinetic theory for the inelastic hard core system 
to the nearly elastic granular gases having the square well potential.
The organization of this paper is as follows.
In the next section, we evaluate the scattering angle for a two-body collision process as a function of the impact parameter and the relative velocity of the colliding pair of particles by solving the Newton equation.
In Sec.\ \ref{sec:kinetic} we extend the kinetic theory for hard core granular gases to the gases having the square well potential to derive the transport coefficients in a set of the hydrodynamic equations.
In Sec.\ \ref{sec:DSMC}, we compare them with those obtained by the DSMC.
In Secs. \ref{sec:Discussion} and \ref{sec:Conclusion}, we discuss and summarize our results, respectively.
In Appendix \ref{sec:collision}, we explain collision geometries for core collisions and grazing collisions to determine the velocity change during collisions in details.
In Appendix \ref{sec:Chapman-Enskog}, we briefly explain the procedure to obtain the transport coefficients by using the Chapman-Enskog theory.
In Appendices \ref{sec:mu2_4} and \ref{sec:Omega}, we calculate the second moment of the collision integral and two Sonine coefficients in terms of the kinetic theory, respectively.
In Appendix \ref{sec:highTexpansion}, we calculate the explicit expressions of the transport coefficients in the high and low temperature limit.
In Appendix \ref{sec:DSMC_algorithm}, we briefly summarize the DSMC algorithm. 
In Appendix \ref{sec:T_escape}, we estimate the critical temperature, at which we cannot ignore the trapping process.


\section{Scattering angle for the square well potential}\label{sec:deflection}
Let us calculate the scattering angle for monodisperse smooth inelastic hard spheres having the square well potential whose mass is $m$ \cite{Hirschfelder1948, Holleran1951, Hirschfelder1954, Resibois, Chapman, Gallagher}.
Here, the hard core potential associated with the square well attractive part for the relative distance $r$ between two spheres is given by
\begin{align}
U(r)=
\begin{cases}
\infty & (r\le d)\\
-\varepsilon & (d<r\le \lambda d)\\
0 & (r>\lambda d)
\end{cases},
\end{align}
where $\varepsilon$ and $\lambda$ are, respectively, the well depth and the well width ratio.
We assume that collisions are inelastic only if particles hit the core ($r=d$) characterized by the restitution coefficient $e$.

\begin{figure}[htbp]
	\begin{center}
		\includegraphics[width=70mm]{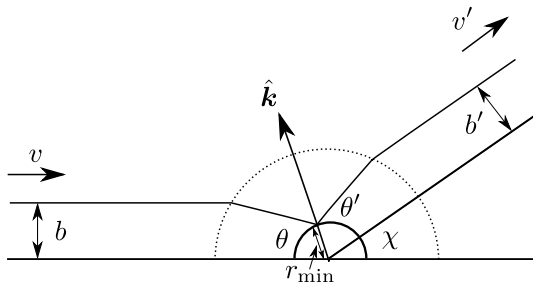}
	\end{center}
	\caption{A schematic view of a collision process. The dotted line represents the outer edge of the attractive potential.}
	\label{fig:collision_process}
\end{figure}

Let us consider a scattering process in which two particles approach from far away with relative velocity $\bm{v}$ and leave with the relative velocity $\bm{v}^\prime$ after the scattering as depicted by Fig.\ \ref{fig:collision_process} in the frame that the target is stationary.
The incident angle $\theta$ between $\bm{v}$ and the normal unit vector $\hat{\bm{k}}$ at the closest distance $r=r_{\rm min}$ between colliding particles is given by
\begin{equation}
\theta = b \int_0^{u_0} \frac{du}{\sqrt{1-b^2 u^2 -\frac{4}{m v^2} U(1/u)}},\label{eq:theta_def}
\end{equation}
where $u\equiv 1/r$. Here, $u_0\equiv 1/r_{\rm min}$ is the smaller one between $1/d$ and the positive solution that the denominator of Eq.~(\ref{eq:theta_def}) is equal to zero \cite{LandauMechanics, GoldsteinMechanics},
and $\hat{\bm{k}}=\bm{r}_{12}/r_{12}$ is a unit vector parallel to $\bm{r}_{12}=\bm{r}_1-\bm{r}_2$ 
with the positions $\bm{r}_1$ and $\bm{r}_2$ for particles 1 and 2, and $r_{12}=|\bm{r}_{12}|$.
We have also introduced the impact parameter $b$ for the incident process.
Because the scattering is inelastic, in general, the impact parameter $b^\prime$ after the scattering and the angle $\theta^\prime$ between $\hat{\bm{k}}$ and $\bm{v}^\prime$ differ from $b$ and $\theta$, respectively (Fig.\ \ref{fig:collision_process}).
Let us consider the case for $b>\lambda d$, where Eq.\ (\ref{eq:theta_def}) reduces to
\begin{equation}
\theta = b \int_0^{1/b} \frac{du}{\sqrt{1-b^2u^2}} = \frac{\pi}{2}
\end{equation}
under the condition $u_0=1/d$.
Because the particles do not collide, $\theta^\prime=\theta$, the scattering angle $\chi$ is given by
\begin{equation}
\chi = \pi -2\theta =0,\quad \sin\frac{\chi}{2}=0.\label{eq:collision3}
\end{equation}

Next, we consider the case for $b\le \lambda d$ in which
Eq.\ (\ref{eq:theta_def}) can be rewritten as
\begin{align}
\theta 
=& b \int_0^{1/\lambda d} \frac{du}{\sqrt{1-b^2u^2}} 
	+ b \int_{1/\lambda d}^{u_0} \frac{du}{\sqrt{1-b^2u^2 + \frac{4\varepsilon}{m v^2}}}\nonumber\\
=& \arcsin \left(\frac{b}{\lambda d}\right)+ b \int_{1/\lambda d}^{u_0} \frac{du}{\sqrt{\nu^2-b^2u^2}},\label{eq:theta_well}
\end{align}
where we have introduced $\nu$ as
\begin{equation}
	\nu\equiv \sqrt{1+\frac{4\varepsilon}{m v^2}}, \label{eq:def_nu}
\end{equation}
and $u_0=\min\left(1/d, \nu/b\right)$ with the introduction of a function $\min(x,y)$ to select the smaller one between $x$ and $y$.
We note that $\nu$ is related to the refractive index \cite{LandauMechanics, GoldsteinMechanics}.
For $b\ge \nu d$, $u_0$ is given by $u_0=\nu/b$ and this collision is called a grazing collision \cite{Hirschfelder1948, Hirschfelder1954, Holleran1951}.
From Eq.\ (\ref{eq:theta_well}), we rewrite $\theta$ as
\begin{equation}
\theta = \frac{\pi}{2}+\arcsin \left(\frac{b}{\lambda d}\right)-\arcsin \left(\frac{b}{\nu \lambda d}\right).\label{eq:theta_grazing}
\end{equation}
Because the particle does not hit the core, $\theta^\prime$ should be equal to $\theta$. Then, the scattering angle $\chi$ is given by
\begin{equation}
\chi =\chi^{(0)}= \pi -2\theta =2\arcsin \left(\frac{b}{\nu \lambda d}\right) - 2\arcsin \left(\frac{b}{\lambda d}\right).\label{eq:deflection_elastic}
\end{equation}
Equation (\ref{eq:deflection_elastic}), thus, can be rewritten as
\begin{align}
\sin \frac{\chi}{2}
=&\sin \left[\arcsin \left(\frac{b}{\nu \lambda d}\right) - \arcsin \left(\frac{b}{\lambda d}\right)\right].\label{eq:collision2}
\end{align}
Note that this collision does not exist for $\lambda< \nu$.

\begin{figure}[htbp]
	\begin{center}
		\includegraphics[width=85mm]{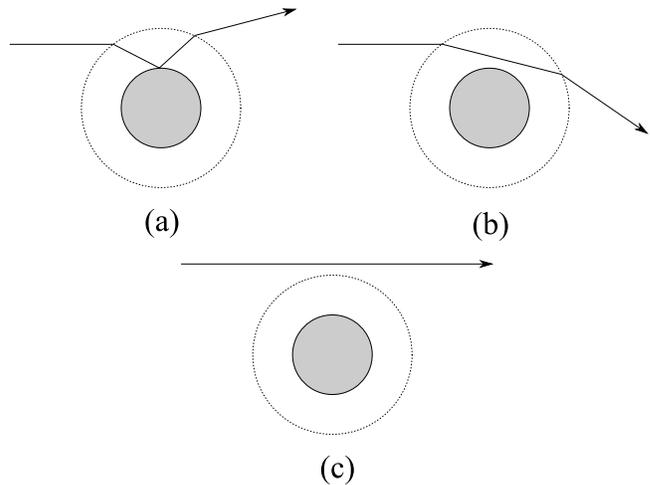}
	\end{center}
	\caption{Schematic views of dynamic processes between two adjacent particles. There exist three types: 
			(a) collisions via the hard core potential (inelastic), 
			(b) grazing collisions (elastic), and (c) no-collisions.}
	\label{fig:collision_type}
\end{figure}

For $b< \nu d$, $u_0$ is given by $u_0=1/d$, and then the particles hit the core of the potential.
From Eq.\ (\ref{eq:theta_well}), we obtain $\theta$:
\begin{equation}
\theta = \arcsin \left(\frac{b}{\lambda d}\right) + \arcsin \left(\frac{b}{\nu d}\right) -\arcsin \left(\frac{b}{\nu \lambda d}\right).
\label{eq:theta_core}
\end{equation}
In this case, the collision is inelastic, and thus, 
$\theta^\prime$ is not equal to $\theta$.
From the conservation of the angular momentum $bv=b^\prime v^\prime$, $\theta^\prime$ is given by 
\begin{widetext}
\begin{align}
\theta^\prime 
=& \arcsin \left(\frac{b^\prime}{\lambda d}\right) + \arcsin \left(\frac{b^\prime}{\nu^\prime d}\right) -\arcsin \left(\frac{b^\prime}{\nu^\prime \lambda d}\right)\nonumber\\
=& \arcsin \left(\frac{b}{\lambda d}\right) 
	+ \arcsin \left(\frac{b}{\nu d}\right) 
	-\arcsin \left(\frac{b}{\nu \lambda d}\right)
+\epsilon \left(\frac{b\nu^2 }{\sqrt{\lambda^2d^2-b^2}} 
	+  \frac{b}{\sqrt{\nu^2d^2b^2}} - \frac{b}{\sqrt{\lambda^2\nu^2d^2-b^2}} \right)\cos^2\Theta
+ \mathcal{O}(\epsilon^2),
\end{align}
\end{widetext}
where we have introduced $\Theta$ as
\begin{equation}
\cos\Theta\equiv \frac{\sqrt{\nu^2d^2-b^2}}{\nu d} \label{eq:cos_Theta}
\end{equation}
(see Appendix \ref{sec:collision} for the derivation) and $\epsilon\equiv 1-e$.
Thus, we obtain the scattering angle $\chi$ as
\begin{equation}
\chi = \pi - \theta - \theta^\prime=\chi^{(0)}+\epsilon\chi^{(1)}+\mathcal{O}(\epsilon^2)\label{eq:chi_bsmall}
\end{equation}
with 
\begin{align}
\chi^{(0)}&= \pi -2\arcsin \left(\frac{b}{\lambda d}\right) \nonumber\\
	&\hspace{1em}- 2\arcsin \left(\frac{b}{\nu d}\right)+ 2\arcsin \left(\frac{b}{\nu \lambda d}\right),\\
\chi^{(1)}&=-\left[\frac{b\nu^2}{\sqrt{\lambda^2d^2-b^2}} 
	+\frac{b}{\sqrt{\nu^2d^2-b^2}} \right.\nonumber\\
	&\hspace{6em}\left.- \frac{b}{\sqrt{\lambda^2\nu^2d^2-b^2}} \right]\cos^2\Theta.
\end{align}
We can rewrite Eq.\ (\ref{eq:chi_bsmall}) as
\begin{align}
\sin\frac{\chi}{2}
&= \sin\frac{\chi^{(0)}}{2}
	+\frac{1}{2}\epsilon \chi^{(1)}\cos\frac{\chi^{(0)}}{2}
	+\mathcal{O}(\epsilon^2).\label{eq:collision1}
\end{align}
These results are consistent with the previous study in the elastic limit ($e\to 1$) \cite{Holleran1951}.
We regard the grazing collision as a combination of (ii) entering and (iii) leaving processes from the well \cite{Holleran1951}.
We ignore the trapping process by the attractive potential in the elastic limit (i.\ e.\ $\epsilon\to 0$) because colliding particles against hard cores have positive energies and the most of rebounding particles have still positive energies.
In other words, if the trapping process is relevant, the inelastic Boltzmann equation is no longer valid.
Thus, through the analysis of the inelastic Boltzmann equation we will discuss whether it can be used even for weakly inelastic cohesive granular gases.
We summarize the above results in Fig.\ \ref{fig:collision_type} and Table \ref{fig:collisions}.

\begin{table}[htbp]
	\caption{Parameters corresponding to Fig.\ \ref{fig:collision_type}.}
	\begin{tabular}{c|c|c|c}
		& (a) hard core & (b) grazing & (c) no-collision\\ 
		& (inelastic) & (elastic) & \\ \hline
		$b$ & $b/d<\min(\nu,\lambda)$ & $\min(\nu,\lambda)\le b/d<\lambda$ & $b/d\ge \lambda$ \rule[-0.5em]{0mm}{1.5em} \\  \hline 
		$\displaystyle \sin\frac{\chi}{2}$ & Eq.(\ref{eq:collision1})& Eq.(\ref{eq:collision2}) &Eq.(\ref{eq:collision3}) \rule[-1em]{0mm}{2.5em}\\ \hline
	\end{tabular}
	\label{fig:collisions}
\end{table}

\section{kinetic theory and hydrodynamic equations}\label{sec:kinetic}
If we consider a dilute and weakly inelastic homogeneous granular gas, we may use the inelastic Boltzmann equation 
\begin{align}
\left(\frac{\partial}{\partial t}+\bm{v}_1\cdot \bm\nabla\right)f(\bm{r},\bm{v}_1,t)=I(f,f),\label{eq:BE}
\end{align}
where $I(f,f)$ is the collision integral
\begin{widetext}
\begin{align}
I(f,f)=& 
\int d\bm{v}_2 \int d\hat{\bm{k}} \Theta(\min(\lambda,\nu)-\tilde{b}) |\bm{v}_{12}\cdot \hat{\bm{k}}|
	\left[\chi_e \sigma(\chi,v_{12}^{\prime\prime})f(\bm{r},\bm{v}_1^{\prime\prime},t)f(\bm{r},\bm{v}_2^{\prime\prime},t) 
	- \sigma(\chi,v_{12})f(\bm{r},\bm{v}_1,t)f(\bm{r},\bm{v}_2,t)\right]\nonumber\\
&+\int d\bm{v}_2 \int d\hat{\bm{k}} \Theta(\tilde{b}-\min(\lambda,\nu)) |\bm{v}_{12}\cdot \hat{\bm{k}}|
	\left[ \sigma(\chi,v_{12}^{\prime\prime})f(\bm{r},\bm{v}_1^{\prime\prime},t)f(\bm{r},\bm{v}_2^{\prime\prime},t) 
	- \sigma(\chi,v_{12})f(\bm{r},\bm{v}_1,t)f(\bm{r},\bm{v}_2,t)\right].\label{eq:collision_term}
\end{align}
\end{widetext}
Here we have introduced the step function $\Theta(x)=1$ for $x>0$ and $\Theta(x)=0$ otherwise.
Here $v_{12}=|\bm{v}_{12}|$ with $\bm{v}_{12}=\bm{v}_1-\bm{v}_2$ with the velocity $\bm{v}_i$ ($i=1,2$) for $i$-th particle, 
$\sigma(\chi, v_{12})$ is the collision cross section between $i$-th and $j$-th particles,
and $\tilde{b}=b/d$ is a dimensionless collision parameter.
The factor $\chi_e$ is related to the Jacobian of the transformation between pre-collisional velocities $\bm{v}_1^{\prime\prime}, \bm{v}_2^{\prime\prime}$ and the velocities after collision $\bm{v}_1, \bm{v}_2$ \cite{Brilliantov, Brilliantov2003, Goldshtein1995, Huthmann2000}.
The first and second terms on the right-hand-side of Eq.\ (\ref{eq:collision_term}) correspond to inelastic and elastic collisions, respectively.
For the sake of later discussion, we explicitly write the relationship between $(\bm{v}_1^{\prime\prime}, \bm{v}_2^{\prime\prime})$ and $(\bm{v}_1,\bm{v}_2)$
\begin{align}
\bm{v}_1 = 
\bm{v}_1^{\prime\prime}+\frac{1}{2}\Delta\bm{v},\quad
\bm{v}_2 = 
\bm{v}_2^{\prime\prime}-\frac{1}{2}\Delta\bm{v},\label{eq:vel_change}
\end{align}
with
\begin{equation}
\Delta \bm{v}=-2\left(1-\frac{1}{2}\epsilon \nu^2 \frac{\cos^2\Theta}{\cos^2\theta}\right)
	(\bm{v}_{12}^{\prime\prime}\cdot \hat{\bm{k}})\hat{\bm{k}} + \mathcal{O}(\epsilon^2)\label{eq:vel_change1}
\end{equation}
for inelastic hard core collisions and
\begin{equation}
\Delta \bm{v}=-2 (\bm{v}_{12}^{\prime\prime}\cdot \hat{\bm{k}})\hat{\bm{k}}\label{eq:vel_change2}
\end{equation}
for elastic grazing collisions (see Appendix \ref{sec:collision} for the derivation).
From Eq.\ (\ref{eq:vel_change1}), the explicit form of the factor $\chi_e$ is given by
\begin{equation}
\chi_e = 1+2\epsilon\nu^2 \frac{\cos^2\Theta}{\cos^2\theta} + \mathcal{O}(\epsilon^2) \label{eq:chi_e}
\end{equation}
for inelastic hard core collisions.
It should be noted that Eq.\ (\ref{eq:chi_e}) is consistent with $1/e^2$ for inelastic hard core potential \cite{Goldshtein1995, Huthmann2000, Brilliantov, Brilliantov2003}, because this can be expanded as $1/e^2=1+2\epsilon +\mathcal{O}(\epsilon^2)$ in the nearly elastic limit and $\nu$ and $\Theta$ reduce to $\nu\to1$ and $\Theta\to \theta$, respectively, in the hard core limit from Eqs.\ (\ref{eq:def_nu}) and (\ref{eq:cos_Theta}).

\subsection{Homogeneous freely cooling}
In this subsection, let us determine the velocity distribution function $f(\bm{v},t)$ in freely cooling granular gases based on the Boltzmann equation (\ref{eq:BE}).
First, we expand the distribution function in terms of Sonine polynomials \cite{Noije1998, Brilliantov, Brilliantov2003, Goldshtein1995, Huthmann2000} as
\begin{equation}
f^{(0)}(\bm{v},t)=f_{\rm M}(V)\left[1+\sum_{{\ell}=1}^\infty a_{\ell} S_{\ell}\left(\frac{mV^2}{2T(t)}\right)\right],\label{eq:zeroth_distribution}
\end{equation}
where $V=|\bm{V}|=|\bm{v}-\bm{U}|$ is the local velocity fluctuation from the flow velocity $\bm{U}(\bm{r},t)$, 
$f_{\rm M}(V)=n(m/2\pi T)^{3/2}\exp(-mV^2/2T)$ is the Maxwellian at the temperature $T$ and the number density $n$, 
and $S_{\ell}(x)\equiv S_{\ell}^{(1/2)}(x)$ is the Sonine polynomial:
\begin{equation}
	S_{\ell}^{(j)}(x)=\sum_{k=0}^{\ell} \frac{(-1)^k \Gamma(j+{\ell}+1)}{\Gamma(j+k+1)({\ell}-k)!k!}x^k
\end{equation}
with the Gamma function $\Gamma(x)$.
The time evolution of the granular temperature, obtained by the product of the Boltzmann equation with $m v_1^2/2$ and integrating over $\bm{v}_1$, is written as
\begin{equation}
	\frac{dT}{dt}=-\zeta^{(0)} T,\label{eq:dTdt}
\end{equation}
where we have introduced the cooling rate for the homogeneous gas
\begin{equation}
	\zeta^{(0)}=\frac{2}{3}nd^2\sqrt{\frac{2T}{m}}{\cal M}_2.\label{eq:zeta_mu2}
\end{equation}
Here, ${\cal M}_2$ is the second moment of the dimensionless collision integral
\begin{align}
	{\cal M}_2 = -\int d\bm{c}_1 c_1^2 \tilde{I}(\tilde{f}^{(0)},\tilde{f}^{(0)}),\label{eq:mu2_def}
\end{align}
where we have introduced the dimensionless velocity $\bm{c}_1=\bm{v}_1/v_T(t)$
with the thermal velocity $v_T(t)=\sqrt{2T(t)/m}$,
the dimensionless collision integral $\tilde{I}(\tilde{f}^{(0)},\tilde{f}^{(0)})=(v_T^2/n^2d^2) I(f^{(0)},f^{(0)})$,
and the dimensionless distribution function $\tilde{f}^{(0)}(\bm{c})=(v_T^3/n)f^{(0)}(\bm{v},t)$.
After some manipulation of Eq.\ (\ref{eq:mu2_def}), ${\cal M}_2$ can be rewritten as \cite{Noije1998,Brilliantov}
\begin{align}
	{\cal M}_2 
	=&-\frac{1}{2}\int d\bm{c}_1 \int d\bm{c}_2 \int d\hat{\bm{k}} 
	|\bm{c}_{12}\cdot \hat{\bm{k}}| \tilde{\sigma}(\chi,c_{12}) \nonumber\\
	&\hspace{3em}\times \tilde{f}^{(0)}(\bm{c}_1) \tilde{f}^{(0)}(\bm{c}_2)
	\Delta [c_1^2+c_2^2]\label{eq:mu2_general}
\end{align}
with $\tilde{\sigma}(\chi,c_{12})=\sigma(\chi,v_{12})/d^2$
and $\phi(c)=\pi^{-3/2}\exp(-c^2)$, 
and $\Delta \psi(\bm{c}_i)\equiv \psi(\bm{c}_i^\prime)-\psi(\bm{c}_i)$.
It should be noted that the density keeps constant and the flow velocity is zero in the homogeneous state.

\subsection{Hydrodynamic equations}

In this subsection, let us derive the transport coefficients which appear in a set of hydrodynamic equations.
Multiplying the Boltzmann equation (\ref{eq:BE}) by $1$, $\bm{v}_1$ and $m v_1^2/2$ and integrating over $\bm{v}_1$,
we obtain the hydrodynamic equations
\begin{align}
&\frac{\partial n}{\partial t} + \bm\nabla \cdot (n\bm{U}) = 0,\label{eq:hydro1}\\
&\frac{\partial \bm{U}}{\partial t} + \bm{U}\cdot \bm\nabla \bm{U} + \frac{1}{mn}\bm\nabla \cdot P=0,\label{eq:hydro2}\\
&\frac{\partial T}{\partial t}+\bm{U}\cdot \bm\nabla T + \frac{2}{3n} \left(P:\bm\nabla \bm{U} 
	+ \bm\nabla \cdot \bm{q}\right) + \zeta T=0,\label{eq:hydro3}
\end{align}
where $n(\bm{r},t)$ is the density field, $\bm{U}(\bm{r},t)$ is the flow velocity, and $T(\bm{r},t)$ is the granular temperature.
The pressure tensor $P$, the heat flux $\bm{q}$, and the cooling rate $\zeta$ are, respectively, defined as
\begin{align}
P_{ij} &\equiv \int d\bm{v} D_{ij}(\bm{V}) f(\bm{r},\bm{v},t)+nT\delta_{ij},\label{eq:pressure}\\
\bm{q} &\equiv \int d\bm{v} \bm{S}(\bm{V}) f(\bm{r},\bm{v},t),\label{eq:heat_flux}\\
\zeta &\equiv -\frac{m}{3nT} \int d\bm{v} v^2 I(f,f),
\end{align}
where $D_{ij}(\bm{V})\equiv m(V_iV_j -V^2\delta_{ij}/3)$ and $\bm{S}(\bm{V})\equiv (mV^2/2-5T/2)\bm{V}$.
We adopt the constitutive equations at the Navier-Stokes order
\begin{align}
P &= p\delta_{ij} -\eta\left(\nabla_i U_j + \nabla_j U_i -\frac{2}{3}\delta_{ij} \bm\nabla \cdot \bm{U}\right),\\
\bm{q} &= -\kappa\bm\nabla T - \mu \bm\nabla n,\label{eq:P_q}
\end{align}
where $p$ is the hydrostatic pressure, $\eta$ is the shear viscosity, 
$\kappa$ is the thermal conductivity, and $\mu$ is the coefficient proportional to the density gradient.
Throughout this paper, we have assumed that the equation of the state $p=nT$ is held because we are interested in the behavior in the dilute limit, though this assumption might not be true if the granular temperature is sufficiently low.

To obtain the transport coefficients, 
we adopt the Chapman-Enskog method \cite{Chapman, Brilliantov, Brilliantov2003}.
Here, we expand the distribution function around Eq.\ (\ref{eq:zeroth_distribution}) as
\begin{equation}
f=f^{(0)} + \delta f^{(1)} + \cdots
\end{equation}
by a small parameter $\delta$ corresponding to the gradients of the fields.
Similarly, the time derivative of the distribution function is expanded as
\begin{equation}
\frac{\partial}{\partial t} 
= \frac{\partial^{(0)}}{\partial t} + \delta \frac{\partial^{(1)}}{\partial t} + \cdots.
\end{equation}
We, thus, rewrite the Boltzmann equation (\ref{eq:BE}) as
\begin{align}
&\left(\frac{\partial^{(0)}}{\partial t} + \delta\frac{\partial^{(1)}}{\partial t} + \cdots + \delta \bm{v}_1 \cdot \bm\nabla\right)
\left(f^{(0)} + \delta f^{(1)} + \cdots\right)\nonumber\\
&=I\left[\left(f^{(0)} + \delta f^{(1)} + \cdots\right),\left(f^{(0)} + \delta f^{(1)} + \cdots\right)\right].\label{eq:Chapman-Enskog}
\end{align}

The equation at the zeroth order of Eq.~(\ref{eq:Chapman-Enskog}) is reduced to
\begin{equation}
\frac{\partial^{(0)}}{\partial t}f^{(0)} = I\left(f^{(0)},f^{(0)}\right).\label{eq:f_0th}
\end{equation}
From Eqs~(\ref{eq:hydro1})--(\ref{eq:hydro3}), the zeroth order hydrodynamic equations are, respectively, given by
\begin{equation}
\frac{\partial^{(0)}}{\partial t}n=0,\quad
\frac{\partial^{(0)}}{\partial t}\bm{U}=0,\quad
\frac{\partial^{(0)}}{\partial t}T=-\zeta^{(0)}T,
\end{equation}
which are equivalent to those obtained in the previous subsection for the homogeneous cooling state.
The zeroth order of the pressure tensor and the heat flux are, respectively, given by
\begin{equation}
P_{ij}^{(0)}=nT\delta_{ij},\quad \bm{q}^{(0)}=0.\label{eq:Pq_0th}
\end{equation}

The first-order Boltzmann equation becomes
\begin{align}
&\frac{\partial^{(0)}}{\partial t}f^{(1)} + \left(\frac{\partial^{(1)}}{\partial t} + \bm{v}_1\cdot \bm\nabla\right)f^{(0)} \nonumber\\
&= I\left(f^{(0)},f^{(1)}\right)+I\left(f^{(1)},f^{(0)}\right).\label{eq:Chapman-Enskog_1st}
\end{align}
The corresponding first-order hydrodynamic equations are, respectively, given by
\begin{align}
\frac{\partial^{(1)}}{\partial t}n &=-\bm\nabla \cdot (n\bm{U}),\nonumber\\
\frac{\partial^{(1)}}{\partial t}\bm{U} &=-\bm{U}\cdot \bm\nabla \bm{U} - \frac{1}{mn}\bm\nabla (nT),\nonumber\\
\frac{\partial^{(1)}}{\partial t}T &=-\bm{U}\cdot \bm\nabla T-\frac{2}{3}T\bm\nabla \cdot \bm{U} - \zeta^{(1)}T,
\end{align}
where the first-order dissipation rate $\zeta^{(1)}$ is defined by
\begin{equation}
\zeta^{(1)}
=-\frac{2m}{3nT} \int d\bm{v} v^2 I\left( f^{(0)},f^{(1)} \right).\label{eq:zeta_1st}
\end{equation}
We note that $\zeta^{(1)}$ becomes zero because of the parity of the integral (\ref{eq:zeta_1st}) 
\cite{Brey1998, Brilliantov, Brilliantov2003}.
We assume that the distribution function $f^{(0)}$ depends on time and space only via its moments:
the density $n$, the average velocity $\bm{U}$ and the temperature $T$ as $f^{(0)}=f^{(0)}[\bm{v}|n,\bm{U},T]$.
Then we can rewrite the first-order equation (\ref{eq:Chapman-Enskog_1st}) as
\begin{align}
&\frac{\partial^{(0)}f^{(1)}}{\partial t} + J^{(1)}\left(f^{(0)},f^{(1)}\right) 
	-\zeta^{(1)}T\frac{\partial f^{(0)}}{\partial T}\nonumber\\
	&=f^{(0)}\left(\bm\nabla \cdot \bm{U} - \bm{V}\cdot \bm\nabla n\right)
		+\frac{\partial f^{(0)}}{\partial T}\left(\frac{2}{3}T\bm\nabla\cdot \bm{U} - \bm{V}\cdot \bm\nabla T\right)\nonumber\\
		&\hspace{3em}+ \frac{\partial f^{(0)}}{\partial \bm{V}}\cdot \left((\bm{V}\cdot \bm\nabla)\bm{U} - \frac{1}{mn}\bm\nabla P\right),\label{eq:Boltzmann_1st_order}
\end{align}
where
\begin{equation}
J^{(1)}\left(f^{(0)},f^{(1)}\right) = -I\left(f^{(0)},f^{(1)}\right)-I\left(f^{(1)},f^{(0)}\right).
\end{equation}
From the form of the first-order equation (\ref{eq:Chapman-Enskog_1st}), 
the solution of this equation is expected to have the form
\begin{equation}
f^{(1)}={\cal \bm{A}}\cdot \bm\nabla \log T + {\cal \bm{B}}\cdot \bm\nabla \log n + {\cal C}_{ij} \nabla_j U_i,\label{eq:1st_solution_assumption}
\end{equation}
where the explicit forms of the coefficients ${\cal \bm{A}}$, ${\cal \bm{B}}$, and ${\cal C}_{ij}$ are given in Appendix \ref{sec:Chapman-Enskog} as Eqs.\ (\ref{eq:cal_A_def}), (\ref{eq:cal_B_def}), and (\ref{eq:cal_C_def}), respectively.
The pressure tensor and the heat flux can be written as
\begin{align}
	P_{ij}^{(1)}=& -\eta \left(\nabla_i U_j + \nabla_j U_i - \frac{2}{3}\delta_{ij}\bm\nabla \cdot \bm{U}\right),\label{eq:P1}\\
	\bm{q}^{(1)}=& -\kappa \bm\nabla T -\mu \bm\nabla n.\label{eq:q1}
\end{align}

Substituting $f=f^{(0)}+f^{(1)}$ and Eq.\ (\ref{eq:P1}) into Eq.\ (\ref{eq:pressure}),
we obtain the differential equation for the shear viscosity $\eta$ with respect to $T$ as
\begin{equation}
	-\zeta^{(0)}T \frac{\partial \eta}{\partial T}
	-\frac{2}{5}nd^2 \sqrt{\frac{2T}{m}}\Omega_\eta^e \eta
	=nT,\label{eq:eta_eq}
\end{equation}
where $\Omega_\eta^e$ is given by
\begin{align}
\Omega^{e}_{\eta} 
	=& \int d\bm{c}_1 \int d\bm{c}_2 \int d\hat{\bm{k}} 
	\tilde{\sigma}(\chi, c_{12})(\bm{c}_{12}\cdot \hat{\bm{k}})\phi(c_1)\phi(c_2)\nonumber\\
	&\times  \left[1+\sum_{{\ell}=1}^\infty a_{\ell} S_{\ell}(c_1^2)\right] \tilde{D}_{ij}(\bm{c}_2)
	\Delta \left[ \tilde{D}_{ij}(\bm{c}_1)+\tilde{D}_{ij}(\bm{c}_2) \right]\label{eq:Omega_eta_general}
\end{align}
with $\tilde{D}_{ij}=D_{ij}/\varepsilon$.
Similarly, substituting Eq.\ (\ref{eq:q1}) into Eq.\ (\ref{eq:heat_flux}), we obtain the differential equations for the thermal conductivity $\kappa$ and the coefficient $\mu$ with respect to $T$ as
\begin{equation}
\frac{\partial}{\partial T} \left(3\zeta^{(0)} \kappa T\right)
	+\frac{4}{5} \kappa nd^2 \sqrt{\frac{2T}{m}}\Omega_\kappa^e
	=-\frac{15}{2}\frac{nT}{m} \left(1+2a_2\right),\label{eq:kappa_eq}
\end{equation}
and
\begin{equation}
-3n \zeta^{(0)} \frac{\partial \mu}{\partial T}-3\kappa \zeta^{(0)} 
	-\frac{4}{5}n^2d^2 \sqrt{\frac{2}{mT}}\Omega_\kappa^e \mu = a_2 \frac{15}{2}\frac{nT}{m},\label{eq:mu_eq}
\end{equation}
respectively, where $\Omega_\kappa^e$ is given by
\begin{align}
\Omega_\kappa^e
	=& \int d\bm{c}_1 \int d\bm{c}_2 \int d\hat{\bm{k}}
	\tilde{\sigma}(\chi, c_{12}) (\bm{c}_{12}\cdot \hat{\bm{k}}) \phi(c_1) \phi(c_2)\nonumber\\
	&\times \left[1+\sum_{{\ell}=1}^\infty a_{\ell} S_{\ell}(c_1^2)\right] 
	\tilde{\bm{S}}(\bm{c}_2) \cdot \Delta \left[ \tilde{\bm{S}}(\bm{c}_1) + \tilde{\bm{S}}(\bm{c}_2) \right]
	\label{eq:Omega_kappa_general}
\end{align}
with $\tilde{\bm{S}}=\bm{S} \sqrt{m/\varepsilon^3}$.
It should be noted that Eqs.\ (\ref{eq:eta_eq}), (\ref{eq:kappa_eq}), and (\ref{eq:mu_eq}) are consistent with those in the previous study in the hard core limit \cite{Brilliantov}.

\subsection{Transport coefficients for the granular gases having the square well potential}\label{sec:TC_SW}

\begin{figure}
	\begin{center}
		\includegraphics[width=85mm]{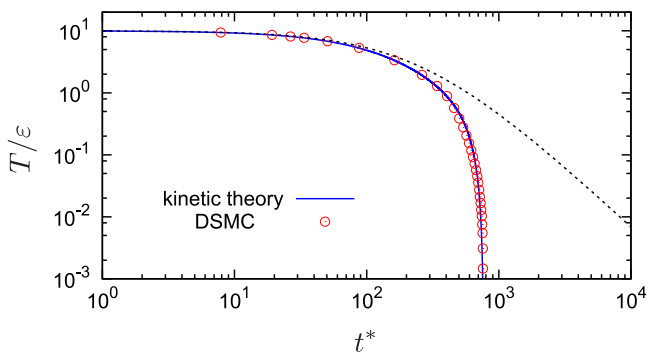}
	\end{center}
	\caption{(Color online) The time evolution of the granular temperature for $nd^3=0.05$, $\lambda=1.5$, and $e=0.99$
			obtained by the kinetic theory (blue solid line) and that by the DSMC (red open circles), 
			where $t^*=t\sqrt{\varepsilon/m}/d$ and the initial temperature is set to be $10\varepsilon$.
			The dotted line represents Haff's law for inelastic hard core spheres in which each particle has the diameter $d$.}
	\label{fig:temp_kin}
\end{figure}

In the previous subsection, we have presented the general framework for the second moment (\ref{eq:mu2_general}) and 
the differential equations of the transport coefficients (\ref{eq:eta_eq}), (\ref{eq:kappa_eq}), and (\ref{eq:mu_eq}) in dilute granular cohesive granular gases without specification of mutual interactions between grains.
In this subsection, let us derive the explicit forms of them for the square well potential outside and the hard core potential inside.
Here, we assume that the zero-th order distribution function can be well reproduced by the truncation up to the third order Sonine polynomials \cite{Noije1998, Brilliantov, Brilliantov2006, Santos2009, Chamorro2013} as
\begin{align}
	\tilde{f}^{(0)}(\bm{c})=\phi(c)\left[1+a_2 S_2(c^2) + a_3 S_3(c^2)\right],\label{eq:f0_a2_a3}
\end{align}
where $a_1$ is automatically zero because the first order moment is absorbed in the definition of the zeroth velocity distribution function.
In this paper, we only consider the elastic limit $\epsilon\to0$.
In addition, the coefficients $a_2$ and $a_3$ can be, respectively, written as the series of $\epsilon$ as shown in Appendix \ref{sec:mu2_4}, 
\begin{equation}
	\begin{cases}
		a_2=a_2^{(0)}+\epsilon a_2^{(1)}+\mathcal{O}(\epsilon^2)\\
		a_3=a_3^{(0)}+\epsilon a_3^{(1)}+\mathcal{O}(\epsilon^2)
	\end{cases},\label{eq:a2a3}
\end{equation}
where the coefficients are given by
\begin{equation}
	a_2^{(0)}=a_3^{(0)}=0,\quad a_2^{(1)}=\frac{N_2}{D},\quad a_3^{(1)}=\frac{N_3}{D}
\end{equation}
with
\begin{widetext}
\begin{align}
N_2
=&2 \int_0^\infty dc_{12} \int_0^{\tilde{b}_{\rm max}} d\tilde{b}\hspace{0.2em} 
	\tilde{b}(\nu^2-\tilde{b}^2)c_{12}^5 (5-c_{12}^2) \exp\left(-\frac{1}{2}c_{12}^2\right) 
	\int_0^\infty dc_{12}^\prime \int_0^\lambda d\tilde{b}^\prime\hspace{0.2em} 
	\tilde{b}^\prime c_{12}^{\prime7}(35-c_{12}^{\prime4})
	\sin^2\chi^{(0)\prime} \exp\left(-\frac{1}{2}c_{12}^{\prime2}\right)\nonumber\\
&- \int_0^\infty dc_{12} \int_0^{\tilde{b}_{\rm max}} d\tilde{b}\hspace{0.2em} 
	\tilde{b}(\nu^2-\tilde{b}^2)c_{12}^5(105-14c_{12}^2-c_{12}^4) \exp\left(-\frac{1}{2}c_{12}^2\right)\nonumber\\
	&\hspace{2em}\times
	 \int_0^\infty dc_{12}^\prime \int_0^\lambda d\tilde{b}^\prime\hspace{0.2em} 
	\tilde{b}^\prime c_{12}^{\prime7}(7-c_{12}^{\prime2}) 
	\sin^2\chi^{(0)\prime} \exp\left(-\frac{1}{2}c_{12}^{\prime2}\right),
\end{align}
\begin{align}
N_3
=&4 \int_0^\infty dc_{12} \int_0^{\tilde{b}_{\rm max}} d\tilde{b}\hspace{0.2em} 
	\tilde{b}(\nu^2-\tilde{b}^2)c_{12}^5(105-14c_{12}^2-c_{12}^4) \exp\left(-\frac{1}{2}c_{12}^2\right)
	\int_0^\infty dc_{12}^\prime \int_0^\lambda d\tilde{b}^\prime\hspace{0.2em} 
	\tilde{b}^\prime c_{12}^{\prime7}
	\sin^2\chi^{(0)\prime} \exp\left(-\frac{1}{2}c_{12}^{\prime2}\right)\nonumber\\
&-8 \int_0^\infty dc_{12} \int_0^{\tilde{b}_{\rm max}} d\tilde{b}\hspace{0.2em} 
	\tilde{b}(\nu^2-\tilde{b}^2)c_{12}^5(5-c_{12}^2) \exp\left(-\frac{1}{2}c_{12}^2\right)
	\int_0^\infty dc_{12}^\prime \int_0^\lambda d\tilde{b}^\prime\hspace{0.2em} 
	\tilde{b}^\prime c_{12}^{\prime7} (7+c_{12}^{\prime2})
	\sin^2\chi^{(0)\prime} \exp\left(-\frac{1}{2}c_{12}^{\prime2}\right),\\
D
=& \int_0^\infty dc_{12} \int_0^\lambda d\tilde{b}\hspace{0.2em} 
	\tilde{b}c_{12}^7 \sin^2\chi^{(0)} \exp\left(-\frac{1}{2}c_{12}^2\right)
	\int_0^\infty dc_{12}^\prime \int_0^\lambda d\tilde{b}^\prime\hspace{0.2em} 
	\tilde{b}^\prime c_{12}^{\prime7}(35-c_{12}^{\prime4}) 
	\sin^2\chi^{(0)\prime} \exp\left(-\frac{1}{2}c_{12}^{\prime2}\right)\nonumber\\
&- \int_0^\infty dc_{12} \int_0^\lambda d\tilde{b}\hspace{0.2em} 
	\tilde{b}c_{12}^7(7-c_{12}^2) \sin^2\chi^{(0)} \exp\left(-\frac{1}{2}c_{12}^2\right)
	\int_0^\infty dc_{12}^\prime \int_0^\lambda d\tilde{b}^\prime\hspace{0.2em} 
	\tilde{b}^\prime c_{12}^{\prime7}(7+c_{12}^{\prime2}) 
	\sin^2\chi^{(0)\prime} \exp\left(-\frac{1}{2}c_{12}^{\prime2}\right).\label{eq:D_a2a3}
\end{align}
\end{widetext}
For simplicity we have introduced the notation $\chi^{(0)\prime}=\chi^{(0)}(\tilde{b}^\prime,c_{12}^\prime)$.
To obtain these expressions, we have ignored the terms proportional to $a_2^2$, $a_3^2$, and $a_2 a_3$ because we are interested in nearly elastic situations.
Therefore, from Eq.\ (\ref{eq:mu2_general}), we obtain
\begin{align}
	{\cal M}_2 =& {\cal M}_2^{(0)}+\epsilon {\cal M}_2^{(1)}
			+ \mathcal{O}(\epsilon^2),\label{eq:mu2_expand}
\end{align}
where
\begin{align}
	{\cal M}_2^{(0)}&=0,\\
	{\cal M}_2^{(1)} &= \sqrt{2\pi} \int_0^\infty dc_{12} \int_0^{\tilde{b}_{\rm max}} d\tilde{b} \hspace{0.2em}
			\tilde{b} (\nu^2-\tilde{b}^2) c_{12}^5 \exp\left(-\frac{1}{2}c_{12}^2\right)
\end{align}
with $\tilde{b}_{\rm max}=\min(\nu(c_{12}),\lambda)$.
Substituting Eqs.\ (\ref{eq:zeta_mu2}) and (\ref{eq:mu2_expand}) into Eq.\ (\ref{eq:dTdt}), we obtain the time evolution of the temperature as the solid line in Fig.\ \ref{fig:temp_kin}, in which the number density, the restitution coefficient, the potential width ratio, and the initial temperature are, respectively, $nd^3=0.05$, $e=0.99$, $\lambda=1.5d$, and $T=10\varepsilon$.
When we start from the temperature much higher than the well-depth, the decreases of the temperature obeys Haff's law for hard core systems in the initial stage \cite{Haff1983}. 
As the temperature approaches the well-depth, the rate of temperature decrease is larger than Haff's law.
A similar result on the crossover from Haff's law to a faster decrease of the temperature has already been reported by Ref.\ \cite{Murphy2015}.

Next, let us calculate the transport coefficients.
Similar to the previous case, with the dropping the contributions from $a_2^2$, $a_3^2$, and $a_2a_3$, the coefficients $\Omega_\eta^e$ and $\Omega_\kappa^e$ defined in Eqs.\ (\ref{eq:Omega_eta_general}) and (\ref{eq:Omega_kappa_general}) are, respectively, given by (see Appendix \ref{sec:Omega} for the derivation)
\begin{align}
\begin{cases}
\Omega_\eta^e=\Omega_\eta^{e(0)}+\epsilon \hspace{0.2em}\Omega_\eta^{e(0)} +\mathcal{O}(\epsilon^2)\\
\Omega_\kappa^e=\Omega_\kappa^{e(0)}+\epsilon \hspace{0.2em}\Omega_\kappa^{e(0)} +\mathcal{O}(\epsilon^2)
\end{cases},\label{eq:Omega_expand}
\end{align}
with
\begin{widetext}
\begin{align}
\Omega_\eta^{e(0)} 
=& -\frac{\sqrt{2\pi}}{4} \int_0^\infty dc_{12} \int_0^\lambda d\tilde{b}\hspace{0.2em}
	\tilde{b} c_{12}^7 \sin^2\chi^{(0)} \exp\left(-\frac{1}{2}c_{12}^2\right),\label{eq:Omega_eta0}\\
\Omega_\eta^{e(1)}
=&-a_2^{(1)}\frac{\sqrt{2\pi}}{128} \int_0^\infty dc_{12} \int_0^\lambda d\tilde{b}\hspace{0.2em}
	\tilde{b}c_{12}^7\left(63-18c_{12}^2+c_{12}^4\right)\sin^2\chi^{(0)} \exp\left(-\frac{1}{2}c_{12}^2\right)\nonumber\\
	&-a_3^{(1)}\frac{\sqrt{2\pi}}{1536} \int_0^\infty dc_{12} \int_0^\lambda d\tilde{b}\hspace{0.2em}
	\tilde{b} c_{12}^7 \left(693-297c_{12}^2+33c_{12}^4-c_{12}^6\right)
	\sin^2\chi^{(0)} \exp\left(-\frac{1}{2}c_{12}^2\right)\nonumber\\
	& -\frac{\sqrt{2\pi}}{4} \int_0^\infty dc_{12} \int_0^\lambda d\tilde{b}\hspace{0.2em}
	\tilde{b} c_{12}^7 \chi^{(1)} \sin 2\chi^{(0)} \exp\left(-\frac{1}{2}c_{12}^2\right)\nonumber\\
	& +\sqrt{2\pi} \int_0^\infty dc_{12} \int_0^{\tilde{b}_{\rm max}} d\tilde{b}\hspace{0.2em}
	\tilde{b} (\nu^2-\tilde{b}^2) c_{12}^7 \left(\frac{2}{3}-\sin^2\frac{\chi^{(0)}}{2}\right)\exp\left(-\frac{1}{2}c_{12}^2\right),
	\label{eq:Omega_eta1}\\
\Omega_\kappa^{e(0)} 
=& -\frac{\sqrt{2\pi}}{4} \int_0^\infty dc_{12} \int_0^\lambda d\tilde{b}\hspace{0.2em}
	\tilde{b} c_{12}^7 \sin^2\chi^{(0)} \exp\left(-\frac{1}{2}c_{12}^2\right),\label{eq:Omega_kappa0}
\end{align}
\begin{align}
\Omega_\kappa^{e(1)} 
=& a_2^{(1)}\frac{\sqrt{2\pi}}{128} \int_0^\infty dc_{12} \int_0^\lambda d\tilde{b}\hspace{0.2em}
	\tilde{b} c_{12}^7 \left(63-18c_{12}^2+c_{12}^4\right) \sin^2\chi^{(0)} \exp\left(-\frac{1}{2}c_{12}^2\right)\nonumber\\
	&+a_3^{(1)} \frac{\sqrt{2\pi}}{1536} \int_0^\infty dc_{12} \int_0^\lambda d\tilde{b}\hspace{0.2em}
	\tilde{b} c_{12}^7 \left(693-297c_{12}^2+33c_{12}^4-c_{12}^6\right) \sin^2\chi^{(0)} \exp\left(-\frac{1}{2}c_{12}^2\right)\nonumber\\
	& -\frac{\sqrt{2\pi}}{4} \int_0^\infty dc_{12} \int_0^\lambda d\tilde{b}\hspace{0.2em}
	\tilde{b} c_{12}^7 \chi^{(1)} \sin 2\chi^{(0)} \exp\left(-\frac{1}{2}c_{12}^2\right)\nonumber\\
	& +\sqrt{2\pi} \int_0^\infty dc_{12} \int_0^{\tilde{b}_{\rm max}} d\tilde{b}\hspace{0.2em}
	\tilde{b} (\nu^2-\tilde{b}^2) c_{12}^7 \cos^2\frac{\chi^{(0)}}{2} \exp\left(-\frac{1}{2}c_{12}^2\right)\nonumber\\
	& +\frac{\sqrt{2\pi}}{8} \int_0^\infty dc_{12} \int_0^{\tilde{b}_{\rm max}} d\tilde{b}\hspace{0.2em}
	\tilde{b} (\nu^2-\tilde{b}^2) c_{12}^5 \left(25-11c_{12}^2\right) \exp\left(-\frac{1}{2}c_{12}^2\right).\label{eq:Omega_kappa1}
\end{align}
\end{widetext}
It should be noted that the zeroth order of these quantities, Eqs.\ (\ref{eq:Omega_eta0}) and (\ref{eq:Omega_kappa0}), are the exactly same as the ones obtained by the previous study \cite{Holleran1951}.

Let us perturbatively solve the differential equation of the shear viscosity (\ref{eq:eta_eq}) with respect to the small parameter $\epsilon$.
We expand the shear viscosity as
\begin{equation}
	\eta=\eta^{(0)} + \epsilon \eta^{(1)} + \mathcal{O}(\epsilon^2).\label{eq:eta_expand}
\end{equation}
From Eqs.\ (\ref{eq:mu2_expand}), (\ref{eq:Omega_expand}), and (\ref{eq:eta_expand}), 
we rewrite the differential equation of the shear viscosity (\ref{eq:eta_eq}) as
\begin{align}
&-\frac{2}{3}nd^2\sqrt{\frac{2T}{m}}\left(\epsilon {\cal M}_2^{(1)}+\cdots\right)
T\frac{\partial}{\partial T}\left(\eta^{(0)} + \epsilon \eta^{(1)}+\cdots \right)\nonumber\\
&-\frac{2}{5}nd^2 \sqrt{\frac{2T}{m}}\left(\Omega_\eta^{e(0)}+\epsilon \hspace{0.2em}\Omega_\eta^{e(0)} +\cdots\right)
\left(\eta^{(0)} + \epsilon \eta^{(1)}+\cdots\right) \nonumber\\
&= nT.
\end{align}
Solving the zeroth and first order of this equation, we obtain
\begin{align}
\eta^{(0)}&=-\frac{5}{2d^2}\sqrt{\frac{mT}{2}}\frac{1}{\Omega_\eta^{e(0)}},\label{eq:eta_0}\\
\eta^{(1)}&=-\left(\frac{\Omega_\eta^{e(1)}}{\Omega_\eta^{e(0)}}
	+\frac{5}{3}\frac{{\cal M}_2^{(1)}T}{\Omega_\eta^{e(0)}}\frac{\partial}{\partial T}\right)\eta^{(0)}.\label{eq:eta_1}
\end{align}
Similarly, the thermal conductivity $\kappa$ and the coefficient $\mu$ are, respectively, given by
\begin{align}
\kappa&=\kappa^{(0)} + \epsilon \kappa^{(1)}+\mathcal{O}(\epsilon^2),\label{eq:kappa_expand}\\
\mu&=\mu^{(0)}+\epsilon \mu^{(1)} +\mathcal{O}(\epsilon^2)\label{eq:mu_expand}
\end{align}
with
\begin{align}
\kappa^{(0)}&=-\frac{75}{16d^2}\sqrt{\frac{2T}{m}}\frac{1}{\Omega_\kappa^{e(0)}},\label{eq:kappa_0}\\
\kappa^{(1)}&=-\frac{\Omega_\kappa^{e(1)}}{\Omega_\kappa^{e(0)}} \kappa^{(0)}
			-\frac{75}{8d^2}\sqrt{\frac{2T}{m}} \frac{a_2^{(1)}}{\Omega_\kappa^{e(0)}}\nonumber\\
			&\hspace{2em}-\frac{5}{2d^2}\frac{1}{\sqrt{T}\Omega_\kappa^{e(0)}} \frac{\partial}{\partial T}
				\left( {\cal M}_2^{(1)} \kappa^{(0)} T^{3/2} \right),\\
\mu^{(0)} &= 0,\label{eq:mu_0}\\
\mu^{(1)} &= -\frac{5}{2n} \frac{{\cal M}_2^{(1)} \kappa^{(0)} T}{\Omega_\kappa^{e(0)}}
			-\frac{75}{8nd^2}\sqrt{\frac{T^3}{2m}} \frac{a_2^{(1)}}{\Omega_\kappa^{e(0)}}.
\end{align}
We note that the zeroth order terms of these transport coefficients, Eqs.\ (\ref{eq:eta_0}) and (\ref{eq:kappa_0}) are identical to those obtained by the previous studies \cite{Holleran1951}.

We obtain the expressions of the transport coefficients as Eqs.\ (\ref{eq:mu2_expand}), (\ref{eq:eta_expand}), (\ref{eq:kappa_expand}), and (\ref{eq:mu_expand}).
The above procedure is not practically useful for the simulation of the hydrodynamic equations because we need to calculate the double integrals at every step.
To reduce the calculation cost, we compare the results with high and low temperature expansions.
From the calculation in Appendix \ref{sec:highTexpansion}, we can obtain the explicit expressions of the dissipation rate and the transport coefficients as in Table \ref{fig:expansion_table}.
As a final remark in this section, we note that our results up to $a_2$ order in Eq.\ (\ref{eq:f0_a2_a3}) are almost identical to those up to $a_3$ in the elastic limit.
This ensures that the expansion around the Maxwellian gives well converged results by Eq.\ (\ref{eq:f0_a2_a3}).

\begin{widetext}
\begin{center}
\begin{table}
	\caption{High temperature expansion of each quantity and low temperature expansion of the second moment up to first order of $\varepsilon/T$ and $\epsilon$.}
	\begin{tabular}{l}\hline\hline
		$\displaystyle
		{\cal M}_2=2\sqrt{2\pi}\epsilon \left(1+\frac{\varepsilon}{T}\right)\quad (T\to \infty),\quad
		{\cal M}_2=2\sqrt{2\pi}\epsilon \left(1+\lambda^2\frac{\varepsilon}{T}\right) \quad (T\to0)$\rule[-1em]{0mm}{2.5em} \\
		$\displaystyle
		\Omega_\eta^e=-4\sqrt{2\pi}\left[1+\epsilon\frac{11}{1280}-\frac{\varepsilon}{T}\frac{\lambda-1}{96}
			\left\{2(15\lambda^4+15\lambda^3+2\lambda^2+2\lambda+2)
			+3\lambda^2(\lambda+1)(5\lambda^2-1)\log \frac{\lambda-1}{\lambda+1}\right\}\right]$,\rule[-1em]{0mm}{2.5em}\\
		$\displaystyle
		\Omega_\kappa^e=-4\sqrt{2\pi}\left[1+\epsilon\frac{1989}{1280}-\frac{\varepsilon}{T}\frac{\lambda-1}{96}
			\left\{2(15\lambda^4+15\lambda^3+2\lambda^2+2\lambda+2)
			+3\lambda^2(\lambda+1)(5\lambda^2-1)\log \frac{\lambda-1}{\lambda+1}\right\}\right]$,\rule[-1em]{0mm}{2.5em}\\
		$\displaystyle
		\eta=\frac{5}{16d^2}\sqrt{\frac{mT}{\pi}}\left[1+\epsilon\frac{1567}{3840}
			+\frac{\varepsilon}{T}\frac{\lambda-1}{96}
			\left\{2(15\lambda^4+15\lambda^3+2\lambda^2+2\lambda+2)
			+3\lambda^2(\lambda+1)(5\lambda^2-1)\log \frac{\lambda-1}{\lambda+1}\right\}\right]$,\rule[-1em]{0mm}{2.5em}\\
		$\displaystyle
		\kappa=\frac{75}{64d^2}\sqrt{\frac{T}{\pi m}}\left[1+\epsilon\frac{539}{1280}
			+\frac{\varepsilon}{T}\frac{\lambda-1}{96}
			\left\{2(15\lambda^4+15\lambda^3+2\lambda^2+2\lambda+2)
			+3\lambda^2(\lambda+1)(5\lambda^2-1)\log \frac{\lambda-1}{\lambda+1}\right\}\right]$,\rule[-1em]{0mm}{2.5em}\\
		$\displaystyle
		\mu=\epsilon \frac{1185}{1024nd^2}\sqrt{\frac{T^3}{\pi m}}$.\rule[-1em]{0mm}{2.5em}\\
	\hline\hline
	\end{tabular}
	\label{fig:expansion_table}
\end{table}
\end{center}
\end{widetext}

\section{comparison with the numerical results}\label{sec:DSMC}

To check the validity of the kinetic theory, 
we compare the transport coefficients derived from the kinetic theory in the previous section 
with those obtained by the DSMC,
which is known as the accurate numerical method to solve the Boltzmann equation \cite{Bird, Alexander1997, Garcia, Poschel}.
We note that stochastic treatment of collisions via DSMC ensures the system uniform, which is suitable to measure the transport coefficients.

\subsection{Cooling coefficient}
\begin{figure}[htbp]
	\begin{center}
		\includegraphics[width=80mm]{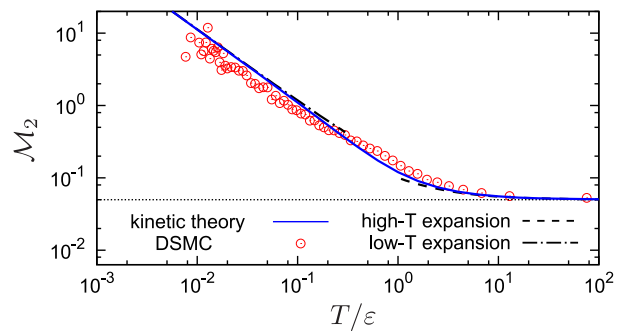}
	\end{center}
	\caption{(Color online) The granular temperature dependence of the second moment ${\cal M}_2$ obtained by the DSMC (red open circles) and that by the kinetic theory up to $a_3$ order (blue solid line), where $T^*$ is the dimensionless temperature defined by $T^*=T/\varepsilon$. The dotted line represents ${\cal M}_2$ for the hard core system with the diameter $d$. The dashed (dot-dashed) line represents ${\cal M}_2$ obtained from the high (low) temperature expansion.}
	\label{fig:mu2_kin}
\end{figure}

In this subsection, we check the time evolution of the granular temperature for homogeneous cooling state and the second moment ${\cal M}_2$.
We prepare monodisperse $N$ particles in a cubic box with the linear system size $L$.
We distribute particles at random as an initial condition,
where the initial velocity distribution obeys Maxwellian with the temperature $T=10\varepsilon$.
Figure \ref{fig:temp_kin} shows the time evolution of the temperature obtained by the DSMC 
and Eq.\ (\ref{eq:dTdt}), in which the number of particles, the system size, the number density, the potential width, and the restitution coefficient are, respectively, $N=6,250$, $L=50d$, $nd^3=0.05$ $\lambda=1.5$, and $e=0.99$.
The time evolution obtained by the kinetic theory fairly agrees with that by the DSMC.
Figure \ref{fig:mu2_kin} shows the comparison of the second moment ${\cal M}_2$ obtained by the kinetic theory with that by the DSMC,
which is also consistent each other, where ${\cal M}_2$ at high temperature limit is identical to that for the hard core system with the diameter $d$.

\subsection{Shear viscosity}
\begin{figure}[htbp]
	\begin{center}
		\includegraphics[width=85mm]{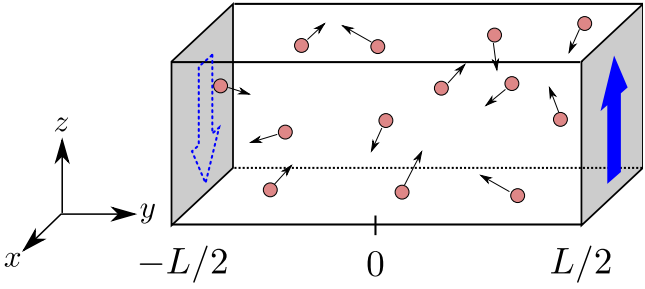}
	\end{center}
	\caption{(Color online) A schematic view of our setup to measure the shear viscosity.
			The walls at $y=L/2$ ($y=-L/2$) move to positive (negative) $z$-direction.}
	\label{fig:eta_setup}
\end{figure}
\begin{figure}[htbp]
	\begin{center}
		\includegraphics[width=85mm]{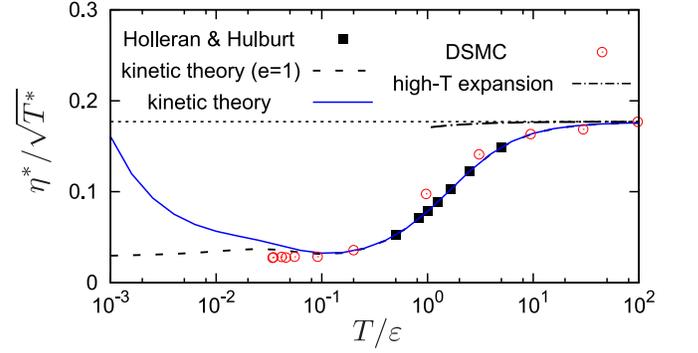}
	\end{center}
	\caption{(Color online) Granular temperature dependence of the shear viscosity obtained by the DSMC (red open circles), that by the elastic kinetic theory (black solid squares in the previous study \cite{Holleran1951} and black dashed line), and that by the kinetic theory (blue solid line), where $\eta^*$ is the dimensionless shear viscosity defined as $\eta^*=\eta d^2/\sqrt{m\varepsilon}$. The dotted line represents the shear viscosity for the hard core system of the diameter $d$. The dot-dashed line represents the shear viscosity obtained from the high temperature expansion.}
	\label{fig:eta_kin}
\end{figure}

%
%
%
Let us compare the result of the shear viscosity by the kinetic theory with that by the DSMC in this subsection.
The particles are distributed at random and the velocity distribution satisfies Maxwellian at the initial condition.
Then, we apply the shear with the aid of the Lees-Edwards walls at $y=\pm L/2$, 
whose $z$-component is $\pm V_{\rm wall}$.
In the initial stage, the energy injection from shear is not balanced with the energy dissipation.
Then, as time goes on, the system reaches a nonequilibrium steady state.
In this stage, we calculate the shear viscosity defined by
\begin{align}
	\eta=- \lim_{t\to \infty} \frac{P_{xy}}{\dot\gamma},
\end{align}
where $\dot\gamma$ is a bulk shear rate defined by the gradient of the flow velocity $U_z$ and $P_{xy}$ can be measured by the DSMC.
To suppress the boundary effects, we measure $\dot\gamma$ in the range $-L/4\le y\le L/4$, that is, $\dot\gamma=(U_z|_{y=L/4}-U_z|_{y=-L/4})/(L/2)$.
Although the Newtonian shear viscosity should be measured by a relaxation process from the initial perturbation for the homogeneous cooling system \cite{Brey1999_2,Bizon1999,Donko2000},
this method is hard to measure the shear viscosity in the low temperature region.
It is also noted that the Newtonian viscosity is known to be identical to the steady state shear viscosity in the elastic limit \cite{Santos2004}, which is the reason why we adopt the above setup.
Figure \ref{fig:eta_kin} shows the comparison of the shear viscosity obtained by the kinetic theory with that by the DSMC, in which the number of particles, the system size, the number density, the potential width, and the restitution coefficient are, respectively, 
$N=10,000$, $L=3,000d$, $nd^3=0.01$ $\lambda=2.5$, and $e=0.99$.
Similar to the case of ${\cal M}_2$, the shear viscosity obtained by the DSMC is identical to that obtained from the kinetic theory for hard-core systems with a particle diameter, $d$, in the high temperature limit.
We cannot measure the shear viscosity for $T\lesssim 10^{-1}\varepsilon$ because the system is heated up by the shear even if we start from a lower temperature.
The first order solution of the kinetic theory with respect to $\epsilon$ also deviates from the zeroth order solution below this temperature, which suggests that the hydrodynamic description is no longer valid in this regime.
This may correspond to the limitation of the inelastic Boltzmann equation, where the trapping processes cannot be ignored even in the elastic limit.

\subsection{Thermal conductivity}
\begin{figure}[htbp]
	\begin{center}
		\includegraphics[width=85mm]{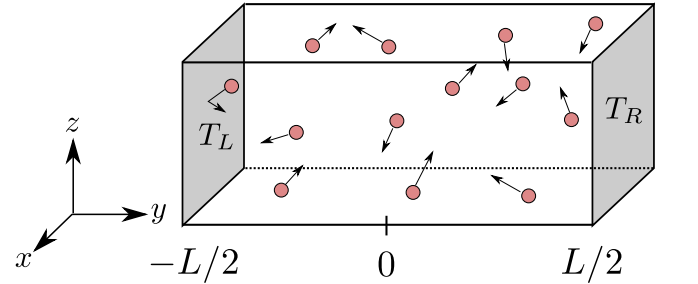}
	\end{center}
	\caption{(Color online) A schematic view of our setup to measure the thermal conductivity.
			The temperature of the left (right) side wall is kept at $T_{\rm L}$ ($T_{\rm R}$).}
	\label{fig:kappa_setup}
\end{figure}
Next, we compare the thermal conductivity obtained by the kinetic theory with that by the DSMC.
Although the heat flux contains the term proportional to the density gradient, we ignore its contribution because the term disappears in the elastic limit $e\to 1$ as in Eq.\ (\ref{eq:mu_0}).
To obtain the thermal conductivity from the DSMC, we solve the heat equation under a confined geometry shown in Fig.\ \ref{fig:kappa_setup}, where the temperature at the left (right) wall at $y=-L/2$ ($y=L/2$) keeps $T_{\rm L}$ ($T_{\rm R}$) \cite{Montanero1994,VegaReyes2009,VegaReyes2010}.
In the steady state, because hydrodynamic variables depend only on $y$, 
the heat equation (\ref{eq:hydro3}) is reduced to
\begin{align}
	\frac{2}{3n}\frac{d}{dy}q_y=\zeta T,\quad q_y=-\kappa \frac{d}{dy}T.
\end{align}
Let us nondimensionalize the quantities using the mass $m$, the system size $L$, and the well depth $\varepsilon$ as
\begin{align}
	&n=\frac{n^*}{L^3},\quad
	y=Ly^*,\quad
	T=\varepsilon T^*,\\
	&p=\frac{\varepsilon}{L^3}p^*,\quad
	{\cal M}_2 = \left(\frac{d}{L}\right){\cal M}_2^*,\quad
	\kappa^\prime = \frac{1}{m^{1/2}L^2}\kappa^{\prime*}.
\end{align}
Thus, we rewrite the heat equation as
\begin{align}
	\frac{d^2}{dy^{*2}}\theta = -3\gamma^2 \theta^{-1/3}\label{eq:heat_nondimensional}
\end{align}
with $\theta=T^{*3/2}$ and $\gamma^2=(1/\sqrt{2})p^{*2}{\cal M}_2^*/\kappa^{\prime *}$.
By multiplying $d\theta/dy^*$ in both sides of Eq.\ (\ref{eq:heat_nondimensional}) 
and integrating the equation from $y^*=0$ to $y^*$, we obtain
\begin{align}
	\frac{d\theta}{dy^*}=\pm\frac{1}{\sqrt{C-9\gamma^2 \theta^{2/3}}},\label{eq:dthetady}
\end{align}
where $C$ is given by $C=\theta_0^{\prime2}+9\gamma^2\theta_0^{2/3}$ 
with $\theta_0=\theta|_{y^*=0}$ and $\theta_0^\prime=d\theta/dy^*|_{y^*=0}$.
Here, we consider the system that the temperature at $y=-L/2$ is lower than that at $y=L/2$,
in which the plus sign is selected in Eq.\ (\ref{eq:dthetady}).
Under this condition, the solution of Eq.\ (\ref{eq:dthetady}) has the following form
\begin{align}
	y^*=\frac{\theta_0^{1/3}}{2\gamma}
	&\left[ -\Theta\sqrt{\beta^2-\Theta^2}
	+ \beta^2 \arctan \left( \frac{\Theta}{\sqrt{\beta^2-\Theta^2}} \right)\right.\nonumber\\
	&\left.+ \sqrt{\beta^2-1}
	-\beta^2 \arctan \left(\frac{1}{\sqrt{\beta^2-1}}\right) \right],\label{eq:T_profile}
\end{align}
where $\beta=\{(\theta^{\prime 2}/9\gamma^2\theta_0^{2/3})+1\}^{1/2}$ and $\Theta=(\theta/\theta_0)^{1/3}$.

\begin{figure}[htbp]
	\begin{center}
		\includegraphics[width=80mm]{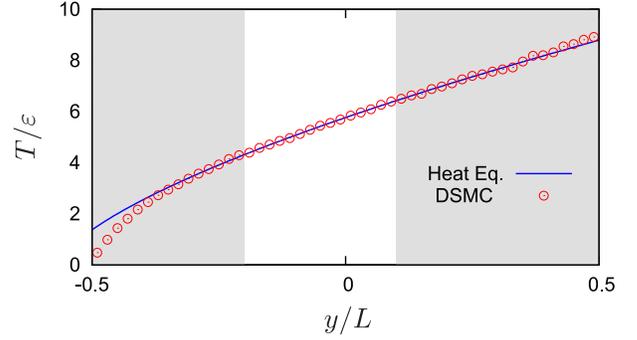}
	\end{center}
	\caption{(Color online) The solution of the heat equation (blue solid line) and the temperature profile obtained by the DSMC (red open circles).
			We choose $\gamma$ to fit the DSMC result in the range $-L/5\le y\le L/10$.}
	\label{fig:T_profile}
\end{figure}

\begin{figure}[htbp]
	\begin{center}
		\includegraphics[width=85mm]{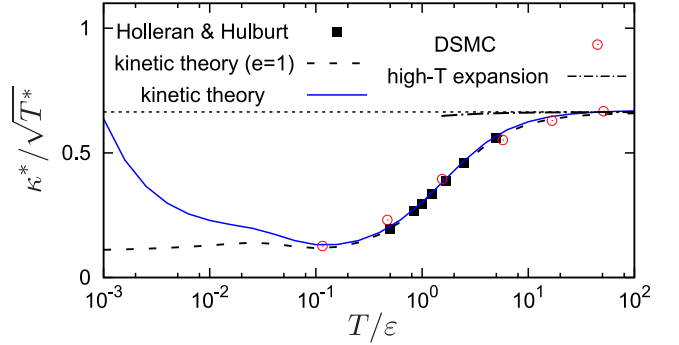}
	\end{center}
	\caption{(Color online) The temperature dependence of the thermal conductivity obtained by the DSMC (red open circles), that by the elastic kinetic theory (black solid squares in the previous study \cite{Holleran1951} and black dashed line), and that by the kinetic theory (blue solid line), where $\kappa^*$ is the dimensionless thermal conductivity defined as $\kappa^*=\kappa d^2\sqrt{m/\varepsilon}$. The dotted line represents the thermal conductivity for the hard core system of the diameter $d$. The dot-dashed line represents the shear viscosity obtained from the high temperature expansion.}
	\label{fig:kappa_prime}
\end{figure}

To obtain $\kappa^\prime$ from the DSMC,
we numerically evaluate $\gamma$ from the comparison of the temperature profile (\ref{eq:T_profile}) with that by the DSMC in the range $-L/5\le y\le L/10$ as in Fig.\ \ref{fig:T_profile}.
It should be noted that we omit the data near the walls to suppress the boundary effects.
Using the estimated $\gamma$ and the simulation results $\theta_0$, $\theta_0^\prime$, and ${\cal M}_2$ in the homogeneous freely cooling, we estimate $\kappa^\prime$ in terms of the DSMC.
Here, the number of particles, the system size, the number density, the potential width, and the restitution coefficient are, respectively, $N=10,000$, $L=3,000d$, $nd^3=0.01$ $\lambda=2.5$, and $e=0.99$.
Figure \ref{fig:kappa_prime} shows the results of the DSMC and the kinetic theory, which is similar to that for $\eta$.
The heat conductivity in the high temperature limit of DSMC is identical to that predicted by the kinetic theory for hard-core systems with a particle diameter $d$.
We note that the profile of the temperature described by Eq.\ (\ref{eq:T_profile}) cannot be achieved for $T\lesssim 10^{-1}\varepsilon$.
Moreover, the perturbative contribution becomes larger than the base value of the perturbation (\ref{eq:kappa_0}) for $T\lesssim 0.1\varepsilon$ as in the case of the viscosity.

\section{Discussion}\label{sec:Discussion}

In this paper, we have obtained the transport coefficients as functions of the granular temperature.
The transport coefficients in high temperature limit are identical to those for the hard core system with the diameter $d$.
Let us consider this reason.
As explained in Sec.\ \ref{sec:deflection}, the collision is inelastic for $b<\min(\nu d,\lambda d)$ while it becomes an elastic grazing collision for $\min(\nu d,\lambda d)<b<\lambda d$.
The value of $\nu= \sqrt{1+4\varepsilon/(mv^2)}$ converges to $1$ in the high temperature limit.
On the other hand, grazing collisions only change the directions of colliding particles and the kinetic energy is kept unchanged.
Therefore, the energy change by collisions in high temperature limit is identical to that for the hard core system of the diameter $d$.

Below $T\simeq 10^{-1}\varepsilon$, the first order solutions of the transport coefficients with respect to $\epsilon$ deviate from the zeroth order solutions.
Moreover, the first order solutions diverge as $T^{-1}$ in the low temperature limit.
This is because $\nu$ diverges as
\begin{equation}
	\nu=\sqrt{1+\frac{2\varepsilon}{T c_{12}^2}} \sim T^{-1/2}
\end{equation}
in the low temperature limit.
This indicates that our hydrodynamic description in terms of the perturbation method is no longer valid for low temperature.

Murphy and Subramaniam \cite{Murphy2015} studied the homogeneous cooling state for a system of particles having an inelastic hard core associated with van der Waals potential.
They obtained that the time evolution of the granular temperature obeys Haff's law in the initial stage and decreases faster as time goes on, then approaches to Haff's law for $e=0$.
They considered that the particles aggregate after the collision when two particles have small kinetic energy with compared to the potential well keeping the potential contribution after the coalescence.
Although we do not consider the aggregation process, the time evolution of the granular temperature in Fig.\ \ref{fig:temp_kin} is similar to their result.

Our theory becomes invalid for $T\lesssim 0.1\varepsilon$ as shown in Figs.\ \ref{fig:eta_kin} and \ref{fig:kappa_prime}.
Let us estimate this critical temperature of coalescence processes from a simple one dimensional collision model.
As explained in Appendix \ref{sec:T_escape}, if the kinetic energy is less than the well depth, the particle cannot escape from the well and be trapped by another particle.
This critical velocity can be estimated as $v_{\rm trap}\simeq \{8(1-e)\varepsilon /m\}^{1/2}$, which leads to the corresponding critical temperature as $T_{\rm trap}=(1/2)mv_{\rm trap}^2\simeq 4(1-e)\varepsilon$.
Using our choice of parameter ($e=0.99$), this temperature becomes $T_{\rm trap}=0.04\varepsilon$, which qualitatively reproduces the lower bound of our theory as shown in Figs.\ \ref{fig:eta_kin} and \ref{fig:kappa_prime}.
Even if we can ignore aggregations of colliding particles, the equation of state $p=nT$ is no longer valid for low temperature regime.
The replacement of the equation of state will be discussed elsewhere.
It should be noted, however, that realistic situations might not be described by Smoluchowski's rate equation as used in Refs.\ \cite{Smoluchowski1916,Chandrasekhar1941,Friedlanger,Krapivsky,Lushnikov1977,Ziff1980,Hendricks1983,Srivastava1983,Hayakawa1987,Krapivsky1996,Spahn2004,Brilliantov 2006,Brilliantov2009,Brilliantov2015} because the biggest cluster may absorb other particles \cite{Takada2014}.
We will study the effects of aggregation processes in the near future.

Here, we have only focused on the dilute system.
To discuss the behavior of a system with finite density is also our future work.


\section{Conclusion}\label{sec:Conclusion}
In this paper, we have developed the kinetic theory for dilute cohesive granular gases having the square well potential to derive the hydrodynamic equations using the Champan-Enskog theory for the inelastic Boltzmann equation.
We have obtained the second moment ${\cal M}_2$ of the collision integral and the transport coefficients for this system.
We have found that they are identical to those for hard core gases at high temperature and the hydrodynamic description is no longer valid at low temperature.
We have also performed DSMC simulation to check the validity of the kinetic theory and found that all results of DSMC are consistent with those obtained by the kinetic theory.

\section{acknowledgments}
The authors thank M.\ Alam for fruitful discussion to initiate this project in the initial stage.
They wish to their express sincere gratitude to A.\ Santos for his continuous encouragement and kind advice.
One of the authors (ST) also thanks M.\ Hattori and S.\ Kosuge for their kind explanation on the DSMC.
The authors appreciate V.\ Garz\'{o} and F.\ Vega Reyes for their suggestive advice on the measurement of the transport coefficients.
Part of this work was performed during the YITP workshops, ``Physics of Glassy and Granular Material" (Grant No.\ YITP-W-13-04) and ``Physics of Granular Flow" (Grant No.\ YITP-T-13-03).
Numerical computation in this work was partially carried out at the Yukawa Institute Computer Facility.
This work is partially supported by Scientific Grant-in-Aid of MEXT, KAKENHI (Nos.\ 25287098 and 16H04025).
This work was also supported by World Premier International Research Center Initiative (WPI), MEXT, Japan.

\appendix

\section{Collision geometry for the square well potential}\label{sec:collision}
\begin{figure}[htbp]
	\begin{center}
		\includegraphics[width=80mm]{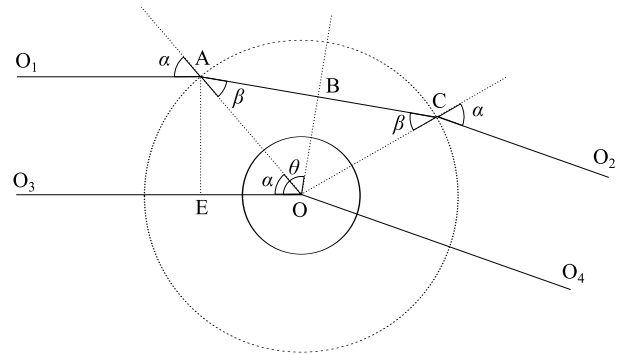}
	\end{center}
	\caption{Collision geometry for a grazing collision. Two particles approach from ${\rm O}_1$ and leave for ${\rm O}_2$.
			The solid and dotted circles represent the hard core (radius $d$) and the outer edge of the potential (radius $\lambda d$), respectively.}
	\label{fig:grazing_collision}
\end{figure}
In this appendix, let us explain the collision geometry scattered by the square well potential.
First, we consider the case for a grazing collision as in Fig.\ \ref{fig:grazing_collision} in the frame that the target is stationary.
Let us consider the process that two particles approach from far away with relative velocity $\bm{v}$ from ${\rm O}_1$.
When the incident particle enters the well at the point A, the relative velocity changes because of the conservation of the energy and the angular momentum, whose speed inside the well is given by $\nu v$.
At the point A, the relative velocity perpendicular to OA is conserved, that is, $v\sin \alpha =\nu v \sin\beta$ is satisfied \cite{LandauMechanics}.
The change of the velocity parallel to OA is given by
\begin{align}
\nu v \cos\beta - v\cos \alpha
&=\nu v \sqrt{1-\frac{1}{\nu^2}\sin^2\alpha} - v\cos\alpha\nonumber\\
&=\left(\sqrt{\nu^2-\sin^2\alpha}-\cos\alpha\right)v,
\end{align}
which means that the velocity change $\Delta \bm{v}_{\rm A}$ at the point A satisfies
\begin{equation}
\Delta \bm{v}_{\rm A}=-\left(\sqrt{\nu^2-\sin^2\alpha}-\cos\alpha\right)v \hat{\bm{r}}_{\rm A} \label{eq:dv_A}
\end{equation}
with the unit vector $\hat{\bm{r}}_{\rm A}=(\cos(\pi-\alpha), \sin(\pi-\alpha))^{\rm T}$ parallel to OA.
We note that the minus sign in Eq.\ (\ref{eq:dv_A}) comes from the fact that the velocity change is opposite direction to $\hat{\bm{r}}_{\rm A}$.

Similarly, the component of the velocity change parallel to OC at the point C is given by $(\cos\alpha-\sqrt{\nu^2-\sin^2\alpha})v$,
which means that the velocity change $\Delta \bm{v}_{\rm C}$ at the point C becomes
\begin{equation}
\Delta \bm{v}_{\rm C}=-\left(\sqrt{\nu^2-\sin^2\alpha}-\cos\alpha\right)v \hat{\bm{r}}_{\rm C} \label{eq:dv_C}
\end{equation}
with the unit vector $\hat{\bm{r}}_{\rm C}=(\cos(\pi-2\theta+\alpha), \sin(\pi-2\theta+\alpha))^{\rm T}$.

From Eqs.\ (\ref{eq:dv_A}) and (\ref{eq:dv_C}), the velocity change $\Delta \bm{v}$ during this grazing collision becomes
\begin{align}
\Delta \bm{v}
	&=\Delta \bm{v}_{\rm A}+\Delta \bm{v}_{\rm C}\nonumber\\
	&=-2\left(\sqrt{\nu^2-\sin^2\alpha}-\cos\alpha\right)\nonumber\\
	&\hspace{2em}\times v \cos(\theta-\alpha)
		\begin{pmatrix}\cos(\pi-\theta)\\ \sin(\pi-\theta)\end{pmatrix}.\label{eq:dv_grazing1}
\end{align}
From Eq.\ (\ref{eq:theta_grazing}) and $\alpha=\arcsin({\rm AE}/{\rm OA})=\arcsin(b/\lambda d)$, 
the following relationships are satisfied:
\begin{align}
\cos(\theta-\alpha)
	=& \cos\left(\frac{\pi}{2}-\arcsin \frac{b}{\nu \lambda d}\right)
	= \frac{b}{\nu\lambda d},\\
\cos\theta
	=& \sin\left(\arcsin \frac{b}{\nu\lambda d} - \arcsin \frac{b}{\lambda d}\right)\nonumber\\
	=&\sin\left(\arcsin\frac{b}{\nu\lambda d}\right) \cos\left(\arcsin\frac{b}{\lambda d}\right)\nonumber\\
		&-\cos\left(\arcsin\frac{b}{\nu\lambda d}\right) \sin\left(\arcsin\frac{b}{\lambda d}\right)\nonumber\\
	=&\frac{b}{\nu\lambda^2d^2}\left(\sqrt{\lambda^2d^2-b^2} - \sqrt{\nu^2\lambda^2d^2-b^2}\right),
\end{align}
and 
\begin{align}
&\sqrt{\nu^2-\sin^2\alpha}-\cos\alpha\nonumber\\
&=  \frac{1}{\lambda d} \left(\sqrt{\nu^2\lambda^2d^2-b^2} -\sqrt{\lambda^2d^2-b^2}\right).
\end{align}
From these equations, we can rewrite Eq.\ (\ref{eq:dv_grazing1}) as
\begin{align}
\Delta \bm{v}
&= 2 v\cos\theta\begin{pmatrix}\cos(\pi-\theta)\\ \sin(\pi-\theta)\end{pmatrix}\nonumber\\
&= -2 v\cos(\pi-\theta)\begin{pmatrix}\cos(\pi-\theta)\\ \sin(\pi-\theta)\end{pmatrix}\nonumber\\
&= -2 (\bm{v}\cdot \hat{\bm{k}})\hat{\bm{k}},
\end{align}
with the unit vector $\hat{\bm{k}}=(\cos(\pi-\theta), \sin(\pi-\theta))^{\rm T}$.

\begin{figure}[htbp]
	\begin{center}
		\includegraphics[width=80mm]{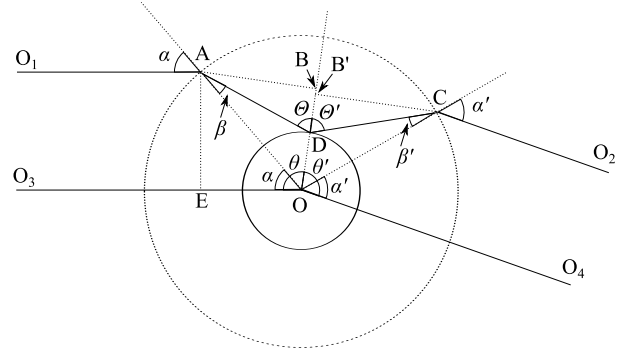}
	\end{center}
	\caption{Collision geometry for a core collision. Two particles approach from ${\rm O}_1$ and leave for ${\rm O}_2$.The solid and dotted lines represent the hard core (radius $d$) and the outer edge of the potential (radius $\lambda d$), respectively.}
	\label{fig:core_collision}
\end{figure}

Next, let us consider the case for a hard core collision as in Fig.\ \ref{fig:core_collision}.
In this case, an inelastic collision takes place at the point D.
To calculate the energy dissipation at the point D, we consider the angle $\Theta$ between the relative velocity of the particle and OB.
From ${\rm AB}=\lambda d \sin(\theta-\alpha)$, ${\rm BD}={\rm OB}-{\rm OD}=(\lambda\cos(\theta-\alpha)-1)d$,
we can write $\Theta$ as
\begin{equation}
\tan \Theta = \frac{\rm AD}{\rm BD}=\frac{\lambda \sin(\theta-\alpha)}{\lambda \cos(\theta-\alpha)-1}.\label{eq:tan_Theta}
\end{equation}
From Eq.\ (\ref{eq:theta_core}), $\cos(\theta-\alpha)$ and $\sin(\theta-\alpha)$ are, respectively, given by
\begin{align}
\cos(\theta-\alpha)
&= \cos\left( \arcsin\frac{b}{\nu d} - \arcsin\frac{b}{\nu\lambda d} \right)\nonumber\\
&= \frac{1}{\nu^2\lambda d^2} \left(\sqrt{\nu^2d^2-b^2}\sqrt{\nu^2\lambda^2d^2-b^2} + b^2\right),\label{eq:cos_theta-alpha}\\
\sin(\theta-\alpha)
&= \sin\left( \arcsin\frac{b}{\nu d} - \arcsin\frac{b}{\nu\lambda d} \right)\nonumber\\
&= \frac{1}{\nu^2\lambda d^2} \left(\sqrt{\nu^2\lambda^2d^2-b^2} - \sqrt{\nu^2d^2-b^2}\right),\label{eq:sin_theta-alpha}
\end{align}
and substituting Eqs.\ (\ref{eq:cos_theta-alpha}) and (\ref{eq:sin_theta-alpha}) into Eq.\ (\ref{eq:tan_Theta}), we obtain
\begin{equation}
\tan\Theta = \frac{b}{\sqrt{\nu^2d^2-b^2}},
\end{equation}
or, equivalently, Eq.\ (\ref{eq:cos_Theta}).
From this, we can calculate the change $\Delta v^2$ after the collision at the point B as
\begin{align}
\Delta v^2 
&=-(1-e^2) \nu^2 v^2 \cos^2\Theta\nonumber\\
&=-(1-e^2) v^2 \left(\nu^2- \frac{b^2}{d^2}\right).
\end{align}
Correspondingly, the change of relative velocity $\Delta \bm{v}$ is given by
\begin{align}
\Delta \bm{v}
&= -\left[(\bm{v}\cdot \hat{\bm{k}}) +\sqrt{(\bm{v}\cdot \hat{\bm{k}})^2-(1-e^2)\nu^2v^2\cos^2\Theta}\right]\hat{\bm{k}}\nonumber\\
&= -2\left[1-\frac{1}{2}\epsilon \nu^2 \frac{\cos^2\Theta}{\cos^2\theta}\right](\bm{v}\cdot \hat{\bm{k}})\hat{\bm{k}}
	+ \mathcal{O}(\epsilon^2),
\end{align}
which reduces to $\Delta \bm{v}=-2(\bm{v}\cdot \hat{\bm{k}})\hat{\bm{k}}$ in the elastic limit.


\section{Chapman-Enskog expansion}\label{sec:Chapman-Enskog}
In this Appendix, let us explain the outline of the Chapman-Enskog theory \cite{Brey1998, Brilliantov}.
As explained in Sec.\ \ref{sec:kinetic}, the zeroth order distribution function, $f^{(0)}$, is determined by Eq.\ (\ref{eq:f_0th}) in the form Eq.\ (\ref{eq:zeroth_distribution}) \cite{Noije1998}.
The first order distribution $f^{(1)}$, satisfies Eq.\ (\ref{eq:Boltzmann_1st_order}), which can be rewritten as
\begin{align}
	&\frac{\partial^{(0)}f^{(1)}}{\partial t} + J^{(1)}\left(f^{(0)},f^{(1)}\right) 
	-\zeta^{(1)}T\frac{\partial f^{(0)}}{\partial T}\nonumber\\
	&=\bm{A}\cdot \bm\nabla \log T + \bm{B}\cdot \bm\nabla \log n + C_{ij}\nabla_j U_i,\label{eq:Boltzmann_1st}
\end{align}
where the coefficients $\bm{A}$, $\bm{B}$, and $C_{ij}$ are, respectively, given by
\begin{align}
\bm{A}(\bm{V})
	&= \frac{1}{2}\bm{V}\frac{\partial}{\partial \bm{V}}\cdot \left(\bm{V} f^{(0)}\right) 
		-\frac{T}{m}\frac{\partial}{\partial \bm{V}}f^{(0)}\nonumber\\
	&=\bm{V} \left[\frac{T}{m}\left(\frac{mV^2}{2T}-1\right) \frac{1}{V}\frac{\partial}{\partial V}+\frac{3}{2}\right]f^{(0)},\label{eq:A_def}\\
\bm{B}(\bm{V})
	&= -\bm{V}f^{(0)} -\frac{T}{m}\frac{\partial}{\partial \bm{V}}f^{(0)}\nonumber\\
	&= -\bm{V}\left(\frac{T}{m} \frac{1}{V}\frac{\partial}{\partial V}+1\right)f^{(0)},\label{eq:B_def}\\
C_{ij}(\bm{V})
	&= \frac{\partial}{\partial V_i}\left(V_j f^{(0)}\right) 
		- \frac{1}{3}\delta_{ij}\frac{\partial}{\partial \bm{V}}\cdot \left(\bm{V} f^{(0)}\right)\nonumber\\
	&=\left(V_i V_j-\frac{1}{3}\delta_{ij}V^2\right)\frac{1}{V}\frac{\partial f^{(0)}}{\partial V}.\label{eq:C_def}
\end{align}
From Eq.\ (\ref{eq:Boltzmann_1st}), $f^{(1)}$ is expected to have the form
\begin{equation}
	f^{(1)}={\cal \bm{A}}\cdot \bm\nabla \log T + {\cal \bm{B}} \cdot \bm\nabla \log n + {\cal C}_{ij}\nabla_j U_i.\label{eq:f1}
\end{equation}
The relationships between the coefficients ${\cal \bm{A}}$, ${\cal \bm{B}}$, ${\cal C}_{ij}$ and $\bm{A}$, $\bm{B}$, $C_{ij}$ are, respectively, obtained by substituting the solution Eq.\ (\ref{eq:f1}) into Eq.\ (\ref{eq:Boltzmann_1st}) as:
\begin{align}
	-T\frac{\partial}{\partial T}\left(\zeta^{(0)}{\cal \bm{A}}\right)
	+J^{(1)}\left(f^{(0)}, {\cal \bm{A}}\right) =& \bm{A},\label{eq:alpha_A}\\
	-\zeta^{(0)}T\frac{\partial {\cal \bm{B}}}{\partial T} - \zeta^{(0)}{\cal \bm{A}}
	+J^{(1)}\left(f^{(0)}, {\cal \bm{B}}\right) =& \bm{B},\label{eq:beta_B}\\
	-\zeta^{(0)}T\frac{\partial {\cal C}_{ij}}{\partial T}
	+J^{(1)}\left(f^{(0)}, {\cal C}_{ij}\right) =& C_{ij},\label{eq:gamma_C}
\end{align}
where we have used $\zeta^{(1)}=0$ because the coefficient $C_{ij}$ is traceless.

Substituting Eq.\ (\ref{eq:f1}) into Eq.\ (\ref{eq:pressure}) with the aid of Eqs.\ (\ref{eq:Pq_0th}) and (\ref{eq:P1}), we obtain
\begin{align}
	&\int d\bm{V} D_{ij}(\bm{V}) {\cal C}_{kl}(\bm{V})\nabla_l U_k \nonumber\\
	&= -\eta \left( \nabla_i U_j + \nabla_j U_i -\frac{2}{3}\delta_{ij} \bm\nabla \cdot \bm{U}\right).
\end{align}
Therefore, the shear viscosity $\eta$ is given by
\begin{equation}
	\eta = -\frac{1}{10}\int d\bm{V} D_{ij}(\bm{V}) {\cal C}_{ji}(\bm{V}).\label{eq:eta_D_C}
\end{equation}
Substituting Eq.\ (\ref{eq:f0_a2_a3}) into Eq.\ (\ref{eq:C_def}), we obtain the explicit form of $C_{ij}(\bm{V})$ as
\begin{align}
&C_{ij}(\bm{V})\nonumber\\
&= -\frac{1}{T} D_{ij}(\bm{V})\left\{ 1+\sum_{\ell} \left[ S_{\ell}(c^2)+S_{{\ell}-1}^{(3/2)}(c^2) \right] \right\}f_{\rm M}(V).
\end{align}
This form and Eq.\ (\ref{eq:gamma_C}) leads to
\begin{align}
	{\cal C}_{ij}(\bm{V})=\frac{{\cal C}_1}{T}D_{ij}(\bm{V})f_{\rm M}(V),\label{eq:cal_C_def}
\end{align}
where ${\cal C}_1$ is a constant.
Substituting Eq.\ (\ref{eq:cal_C_def}) into Eq.\ (\ref{eq:eta_D_C}), we obtain ${\cal C}_1=-\eta/(nT)$.

Similarly, substituting $f^{(1)}$ into Eq.\ (\ref{eq:heat_flux}) with the aid of Eqs.\ (\ref{eq:Pq_0th}) and (\ref{eq:q1}), we obtain
\begin{align}
	\left\{\frac{1}{T} \int d\bm{V} S_i(\bm{V}) {\cal A}_j (\bm{V}) \right\} \nabla_j T =& -\kappa \nabla_i T,\\
	\left\{\frac{1}{n} \int d\bm{V} S_i(\bm{V}) {\cal B}_j (\bm{V}) \right\} \nabla_j n =& -\mu \nabla_i n.
\end{align}
Therefore, we, respectively, obtain the thermal conductivity and the coefficient $\mu$ as
\begin{align}
	\kappa =& -\frac{1}{3T} \int d\bm{V} \bm{S}(\bm{V}) \cdot {\cal \bm{A}} (\bm{V}),\label{eq:kappa_S_A}\\
	\mu=& -\frac{1}{3n} \int d\bm{V} \bm{S}(\bm{V}) \cdot {\cal \bm{B}} (\bm{V}).\label{eq:mu_S_B}
\end{align}
Substituting Eq.\ (\ref{eq:f0_a2_a3}) into Eqs.\ (\ref{eq:A_def}) and (\ref{eq:B_def}), we obtain the explicit forms of $\bm{A}(\bm{V})$ and $\bm{B}(\bm{V})$ as
\begin{widetext}
\begin{align}
	\bm{A}(\bm{V})=& \bm{V}\left\{ S_1^{(3/2)}(c^2)\left[ 1+a_2\left(S_2^{(3/2)}(c^2)-\frac{3}{2}\right) \right] 
				+ \sum_{{\ell}=3}^\infty a_{\ell} \left[ S_1^{(3/2)}(c^2) S_{\ell}(c^2)+(1-c^2)S_{{\ell}-1}^{(3/2)}(c^2) \right]
					 \right\}f_{\rm M}(V),\\
	\bm{B}(\bm{V})=& \sum_{\ell} a_{\ell} \bm{V} S_{{\ell}-1}^{(3/2)}(c^2) f_{\rm M}(V).
\end{align}
\end{widetext}
Equations (\ref{eq:alpha_A}) and (\ref{eq:beta_B}) leads to
\begin{align}
	{\cal A}=& -\frac{{\cal A}_1}{T}\bm{S}(\bm{V})f_{\rm M}(V),\label{eq:cal_A_def}\\
	{\cal B}=& -\frac{{\cal B}_1}{T}\bm{S}(\bm{V})f_{\rm M}(V),\label{eq:cal_B_def}
\end{align}
where ${\cal A}_1$ and ${\cal B}_1$ are constants.
Substituting Eqs.\ (\ref{eq:A_def}) and (\ref{eq:B_def}) into Eq.\ (\ref{eq:kappa_S_A}) and (\ref{eq:mu_S_B}), respectively, and integrating over $\bm{V}$, we obtain ${\cal A}_1=2m\kappa/5nT$ and ${\cal B}_1=2m\mu/5T^2$.

Let us determine the explicit forms of the transport coefficients.
Multiplying Eq.\ (\ref{eq:gamma_C}) by $D_{ij}(\bm{V}_1)$ and integrate over $\bm{V}_1$, we obtain
\begin{align}
	&10\zeta^{(0)} T \frac{\partial \eta}{\partial T}
	+ \int d\bm{V}_1 D_{ij}(\bm{V}_1) J^{(1)}\left(f^{(0)}, {\cal C}_{ij}\right)\nonumber\\
	&= \int d\bm{V}_1 D_{ij}(\bm{V}_1) C_{ij}(\bm{V}_1).\label{eq:eta_eq1}
\end{align}
The second term on the left-hand-side of Eq.\ (\ref{eq:eta_eq1}) is written as
\begin{align}
	\int d\bm{V}_1 D_{ij}(\bm{V}) J^{(1)}\left(f^{(0)}, {\cal C}_{ij}\right)
	=&4\eta n d^2 \sqrt{\frac{2T}{m}} \Omega_\eta^e,
\end{align} 
where $\Omega_\eta^e$ is defined as Eq.\ (\ref{eq:Omega_eta_general}).
Similarly, the right-hand-side of Eq.\ (\ref{eq:eta_eq1}) satisfies
\begin{equation}
	\int d\bm{V}_1 D_{ij}(\bm{V}) C_{ij}(\bm{V}_1)=10nT.
\end{equation}
Therefore, Eq.\ (\ref{eq:eta_eq1}) is reduced to Eq.\ (\ref{eq:eta_eq}).
The perturbative solution of Eq.\ (\ref{eq:eta_eq}) with respect to the small inelasticity is given by Eq.\ (\ref{eq:eta_expand}).

Similarly, we derive the differential equation for the thermal conductivity $\kappa$.
Multiplying Eq.\ (\ref{eq:alpha_A}) by $\bm{S}(\bm{V}_1)/T$ and integrating over $\bm{V}_1$, we obtain
\begin{align}
	&\frac{\partial}{\partial T} \left(3\zeta^{(0)} \kappa T\right)
	+\frac{1}{T}\int d\bm{V}_1 \bm{S}(\bm{V}_1) J^{(1)}\left(f^{(0)},{\cal \bm{A}}\right)\nonumber\\
	&= \frac{1}{T} \int d\bm{V}_1 \bm{S}(\bm{V}_1) \cdot \bm{A}(\bm{V}_1).\label{eq:kappa_eq1}
\end{align}
The second term on the left-hand-side of Eq.\ (\ref{eq:kappa_eq1}) is written as
\begin{equation}
	\frac{1}{T}\int d\bm{V}_1 \bm{S}(\bm{V}_1) J^{(1)}\left(f^{(0)},{\cal \bm{A}}\right)
	=\frac{4}{5}\kappa nd^2 \sqrt{\frac{2T}{m}}\Omega_\kappa^e,
\end{equation}
where $\Omega_\kappa^e$ is given by Eq.\ (\ref{eq:Omega_kappa_general}).
The right-hand-side on Eq.\ (\ref{eq:kappa_eq1}) satisfies
\begin{equation}
	\frac{1}{T} \int d\bm{V}_1 \bm{S}(\bm{V}_1) \cdot \bm{A}(\bm{V}_1)
	= -\frac{15}{2}\frac{nT}{m}\left(1+2a_2\right).
\end{equation}
It should be noted that terms proportional to $a_n$ ($n\ge 3$) vanish due to the orthogonality of the Sonine polynomials.
Therefore, Eq.\ (\ref{eq:kappa_eq1}) is reduced to Eq.\ (\ref{eq:kappa_eq}).
The solution of Eq.\ (\ref{eq:kappa_eq}) is given by Eq.\ (\ref{eq:kappa_expand}).

Similarly, multiplying Eq.\ (\ref{eq:beta_B}) by $\bm{S}(\bm{V}_1)/T$ and integrating over $\bm{V}_1$, the coefficient $\mu$ is given by Eq.\ (\ref{eq:mu_expand}).

\begin{widetext}
\section{Determination of $a_2$ and $a_3$}\label{sec:mu2_4}
In this appendix, we determine the coefficients $a_2$ and $a_3$ using the moments of the dimensionless collision integrals \cite{Brilliantov2006, Santos2009, Chamorro2013}.
It is useful to introduce the basic integral \cite{Brilliantov} 
\begin{align}
J_{k,l,m,n,p,\alpha} \equiv 
	&\int d\bm{C} \int d\bm{c}_{12} \int d\hat{\bm{k}} \tilde{\sigma}
		(\chi,c_{12},\xi) |\bm{c}_{12}\cdot \hat{\bm{k}}|^{1+\alpha} \phi(C) \phi(c_{12})
	C^k c_{12}^l (\bm{C}\cdot \bm{c}_{12})^m (\bm{C}\cdot \hat{\bm{k}})^n (\bm{c}_{12}\cdot \hat{\bm{k}})^p,
\end{align}
with $\bm{C}=(\bm{c}_1+\bm{c}_2)/2$.
This is rewritten as
\begin{align}
J_{k,l,m,n,p,\alpha} 
=&2^{-(k+m+n-1)/2} \Gamma\left(\frac{k+m+n+3}{2}\right)\pi^{-1/2}
	\sum_{j=0}^n \dbinom{n}{j} \left[1+(-1)^j\right] \frac{\Gamma\left(\frac{1+j}{2}\right)}{\Gamma\left(\frac{2+j}{2}\right)}
	\int_0^{\pi}d\Theta \sin^{j+1}\Theta \cos^{m+n-j}\Theta\nonumber\\
	&\times \int_0^\infty dc_{12} \int_0^\infty d\tilde{b} \hspace{0.2em}\tilde{b}c_{12}^{l+m+p+\alpha+3}
		\sin^{n+p-j}\frac{\chi}{2} \left|\sin\frac{\chi}{2}\right|^\alpha \cos^j \frac{\chi}{2} \exp\left(-\frac{1}{2}c_{12}^2\right).\label{eq:def_J}
\end{align}
For $\alpha=0$ and $n=0$, $1$ and $2$, Eq.\ (\ref{eq:def_J}) reduces to
\begin{align}
J_{k,l,m,0,p,0}
=& \frac{2^{-(k+m-3)/2}}{m+1}\left[1+(-1)^m\right] \Gamma \left(\frac{k+m+3}{2}\right)
	 \int_0^{\infty} dc_{12} \int_0^{\infty} d\tilde{b} 
	\hspace{0.2em}\tilde{b} c_{12}^{l+m+p+3}\sin^{p} \frac{\chi}{2} \exp\left(-\frac{1}{2}c_{12}^2\right),\\
J_{k,l,m,1,p,0}
=& \frac{2^{-(k+m-2)/2}}{m+2}\left[1-(-1)^m\right] \Gamma \left(\frac{k+m+4}{2}\right)
	\int_0^{\infty} dc_{12} \int_0^{\infty} d\tilde{b} 
	\hspace{0.2em}\tilde{b} c_{12}^{l+m+p+3}\sin^{p+1} \frac{\chi}{2} \exp\left(-\frac{1}{2}c_{12}^2\right),\\
J_{k,l,m,2,p,0}
=& \frac{2^{-(k+m-1)/2}}{(m+1)(m+3)}\left[1+(-1)^m\right] \Gamma \left(\frac{k+m+5}{2}\right)\nonumber\\
	&\times \int_0^{\infty} dc_{12} \int_0^{\infty} d\tilde{b} 
	\hspace{0.2em}\tilde{b} c_{12}^{l+m+p+3}\sin^p \frac{\chi}{2} \left(1+m\sin^2 \frac{\chi}{2}\right) 
	\exp\left(-\frac{1}{2}c_{12}^2\right),
\end{align}
respectively.
These integrals recover the previous results in the hard core limit \cite{Brilliantov}.
In this paper, we only consider the nearly elastic case $1-e\ll 1$.
We assume that the coefficients $a_2$ and $a_3$ are proportional to $1-e$.
When we use the truncated distribution function Eq.\ (\ref{eq:f0_a2_a3}),
we rewrite the $n$-th moment ${\cal M}_p= -\int d\bm{c}_1 c_1^p \tilde{I}(\tilde{f}^{(0)},\tilde{f}^{(0)})$ ($p\in \mathbb{N}$) as
\begin{align}
{\cal M}_p
=& -\frac{1}{2}
	\int d\bm{C} d\bm{c}_{12} d\hat{\bm{k}} 
	\tilde{\sigma}(\chi,c_{12},\xi)|\bm{c}_{12}\cdot \hat{\bm{k}}|
	\phi(c_1)\phi(c_2)  (\bm{c}_{12}\cdot \hat{\bm{k}})^2 \nonumber\\
	&\hspace{3em}\times\left[1+a_2 (S_2(c_1^2)+S_2(c_2^2))+a_3 (S_3(c_1^2)+S_3(c_2^2))\right]\Delta\left(c_1^p +c_2^p\right),
\end{align}
where, we have ignored the terms proportional to $a_2^2$, $a_3^2$, and $a_2a_3$, because they are the order of $(1-e)^2$.
The explicit forms of $\Delta(c_1^p+c_2^p)$ for $p=2$, $4$, and $6$ are, respectively, given by
\begin{align}
\Delta(c_1^2+c_2^2)
=& -\epsilon \Theta(\tilde{b}_{\rm max}-\tilde{b}) \nu^2\frac{\cos^2\Theta}{\cos^2\theta} 
	(\bm{c}_{12}\cdot \hat{\bm{k}})^2+\mathcal{O}(\epsilon^2),\\
\Delta(c_1^4+c_2^4)
=&-8(\bm{C}\cdot \bm{c}_{12})(\bm{C}\cdot \hat{\bm{k}})(\bm{c}_{12}\cdot \hat{\bm{k}})
	+8(\bm{C}\cdot \hat{\bm{k}})^2 (\bm{c}_{12}\cdot \hat{\bm{k}})^2\nonumber\\
&+ \epsilon \hspace{0.2em}\Theta(\tilde{b}_{\rm max}-\tilde{b}) \nu^2\frac{\cos^2\Theta}{\cos^2\theta} \nonumber\\
	&\hspace{1em}\times\left[
	-2C^2(\bm{c}_{12}\cdot \hat{\bm{k}})^2
	-\frac{1}{2}c_{12}^2(\bm{c}_{12}\cdot \hat{\bm{k}})^2
	+4(\bm{C}\cdot \bm{c}_{12}) (\bm{C}\cdot \hat{\bm{k}}) (\bm{c}_{12}\cdot \hat{\bm{k}})
	-8(\bm{C}\cdot \hat{\bm{k}})^2 (\bm{c}_{12}\cdot \hat{\bm{k}})^2 \right]
	+\mathcal{O}(\epsilon^2),\\
\Delta(c_1^6+c_2^6)
=&-24C^2 (\bm{C}\cdot \bm{c}_{12})(\bm{C}\cdot \hat{\bm{k}})(\bm{c}_{12}\cdot \hat{\bm{k}})
	+24C^2 (\bm{C}\cdot \hat{\bm{k}})^2(\bm{c}_{12}\cdot \hat{\bm{k}})^2\nonumber\\
	&-6 c_{12}^2 (\bm{C}\cdot \bm{c}_{12})(\bm{C}\cdot \hat{\bm{k}})(\bm{c}_{12}\cdot \hat{\bm{k}})
	+6 c_{12}^2 (\bm{C}\cdot \hat{\bm{k}})^2(\bm{c}_{12}\cdot \hat{\bm{k}})^2\nonumber\\
&+\epsilon \hspace{0.2em}\Theta(\tilde{b}_{\rm max}-\tilde{b})\nu^2\frac{\cos^2\Theta}{\cos^2\theta}\nonumber\\
	&\hspace{1em}\times \left[
	3C^4(\bm{c}_{12}\cdot \hat{\bm{k}})^2
	+\frac{3}{2}C^2 c_{12}^2(\bm{c}_{12}\cdot \hat{\bm{k}})^2
	-12C^2 (\bm{C}\cdot \bm{c}_{12})(\bm{C}\cdot \hat{\bm{k}})(\bm{c}_{12}\cdot \hat{\bm{k}})
	+24C^2(\bm{C}\cdot \hat{\bm{k}})^2(\bm{c}_{12}\cdot \hat{\bm{k}})^2\right.\nonumber\\
&\left.\hspace{2em}
	+\frac{3}{16} c_{12}^4(\bm{c}_{12}\cdot \hat{\bm{k}})^2
	-3 c_{12}^2 (\bm{C}\cdot \bm{c}_{12})(\bm{C}\cdot \hat{\bm{k}})(\bm{c}_{12}\cdot \hat{\bm{k}})
	+6 c_{12}^2(\bm{C}\cdot \hat{\bm{k}})^2(\bm{c}_{12}\cdot \hat{\bm{k}})^2
	+3  (\bm{C}\cdot \bm{c}_{12})^2(\bm{c}_{12}\cdot \hat{\bm{k}})^2\right.\nonumber\\
&\left.\hspace{2em}
	-12 (\bm{C}\cdot \bm{c}_{12})(\bm{C}\cdot \hat{\bm{k}})(\bm{c}_{12}\cdot \hat{\bm{k}})^3
	+12(\bm{C}\cdot \hat{\bm{k}})^2(\bm{c}_{12}\cdot \hat{\bm{k}})^4\right]
	+\mathcal{O}(\epsilon^2).
\end{align}
Then, we explicitly write ${\cal M}_2$, ${\cal M}_4$, and ${\cal M}_6$ as
\begin{align}
\begin{cases}
{\cal M}_2 = \sqrt{2\pi}\left(S_1 + a_2S_2 + a_3 S_3\right),\\
{\cal M}_4 = \sqrt{2\pi}\left(T_1 + a_2T_2 + a_3 T_3\right),\\
{\cal M}_6 = \sqrt{2\pi}\left(D_1 + a_2D_2 + a_3 D_3\right),
\end{cases}\label{eq:mu246}
\end{align}
where
\begin{align}
S_1=&\epsilon \int_0^\infty dc_{12} \int_0^{\tilde{b}_{\rm max}}d\tilde{b}\hspace{0.2em} 
	\tilde{b}(\nu^2-\tilde{b}^2)c_{12}^5 \exp\left(-\frac{1}{2}c_{12}^2\right)
	+\mathcal{O}(\epsilon^2),\\
S_2=&\epsilon \frac{1}{16} \int_0^\infty dc_{12} \int_0^{\tilde{b}_{\rm max}}d\tilde{b}\hspace{0.2em} 
	\tilde{b}(\nu^2-\tilde{b}^2)c_{12}^5 (15-10c_{12}^2+c_{12}^4) \exp\left(-\frac{1}{2}c_{12}^2\right)
	+\mathcal{O}(\epsilon^2),\\
S_3=&\epsilon \frac{1}{192} \int_0^\infty dc_{12} \int_0^{\tilde{b}_{\rm max}}d\tilde{b}\hspace{0.2em} 
	\tilde{b}(\nu^2-\tilde{b}^2)c_{12}^5(105-105c_{12}^2+21c_{12}^4-c_{12}^6) \exp\left(-\frac{1}{2}c_{12}^2\right)
	+\mathcal{O}(\epsilon^2),\\
T_1=&\epsilon \frac{1}{2} \int_0^\infty dc_{12} \int_0^{\tilde{b}_{\rm max}}d\tilde{b}\hspace{0.2em} 
	\tilde{b}(\nu^2-\tilde{b}^2)c_{12}^5 (5+c_{12}^2) \exp\left(-\frac{1}{2}c_{12}^2\right)
	+\mathcal{O}(\epsilon^2),\\
T_2=&\frac{1}{4}\int_0^\infty dc_{12} \int_0^\lambda d\tilde{b}\hspace{0.2em} 
	\tilde{b}c_{12}^7 \sin^2\chi^{(0)} \exp\left(-\frac{1}{2}c_{12}^2\right)\nonumber\\
	&+\epsilon \left[
		\frac{1}{32}\int_0^\infty dc_{12} \int_0^{\tilde{b}_{\rm max}}d\tilde{b}\hspace{0.2em} 
		\tilde{b}(\nu^2-\tilde{b}^2)c_{12}^5 (-25-23c_{12}^2-5c_{12}^4+c_{12}^6)
		\exp\left(-\frac{1}{2}c_{12}^2\right)\right.\nonumber\\
	&\hspace{3em}\left.+\int_0^\infty dc_{12} \int_0^{\tilde{b}_{\rm max}}d\tilde{b}\hspace{0.2em} 
		\tilde{b}(\nu^2-\tilde{b}^2)c_{12}^7 \sin^2\frac{\chi^{(0)}}{2}\exp\left(-\frac{1}{2}c_{12}^2\right)\right.\nonumber\\
	&\hspace{3em}\left.+\frac{1}{4}\int_0^\infty dc_{12} \int_0^\lambda d\tilde{b}\hspace{0.2em} 
		\tilde{b}c_{12}^7 \chi^{(1)}\sin^2 2\chi^{(1)} \exp\left(-\frac{1}{2}c_{12}^2\right)\right]
	+\mathcal{O}(\epsilon^2),\\
T_3=&\frac{1}{16}\int_0^\infty dc_{12} \int_0^\lambda d\tilde{b}\hspace{0.2em} 
	\tilde{b}c_{12}^7 (7-c_{12}^2)\sin^2\chi^{(0)} \exp\left(-\frac{1}{2}c_{12}^2\right)\nonumber\\
	&+\epsilon \left[
		\frac{1}{384}\int_0^\infty dc_{12} \int_0^{\tilde{b}_{\rm max}}d\tilde{b}\hspace{0.2em} 
		\tilde{b}(\nu^2-\tilde{b}^2)c_{12}^5 (-525-168c_{12}^2-6c_{12}^4+16c_{12}^6-c_{12}^8)
		\exp\left(-\frac{1}{2}c_{12}^2\right)\right.\nonumber\\
	&\hspace{3em}\left.+\frac{1}{4}\int_0^\infty dc_{12} \int_0^{\tilde{b}_{\rm max}}d\tilde{b}\hspace{0.2em} 
		\tilde{b}(\nu^2-\tilde{b}^2)c_{12}^7 (7-c_{12}^2)\sin^2\frac{\chi^{(0)}}{2}\exp\left(-\frac{1}{2}c_{12}^2\right)\right.\nonumber\\
	&\hspace{3em}\left.+\frac{1}{16}\int_0^\infty dc_{12} \int_0^\lambda d\tilde{b}\hspace{0.2em} 
		\tilde{b}c_{12}^7 (7-c_{12}^2)\chi^{(1)}\sin^2 2\chi^{(1)} \exp\left(-\frac{1}{2}c_{12}^2\right)\right]
	+\mathcal{O}(\epsilon^2),\\
D_1=&\epsilon \frac{3}{16} \int_0^\infty dc_{12} \int_0^{\tilde{b}_{\rm max}}d\tilde{b}\hspace{0.2em} 
	\tilde{b}(\nu^2-\tilde{b}^2)c_{12}^5 (35+14c_{12}^2+c_{12}^4) \exp\left(-\frac{1}{2}c_{12}^2\right)
	+\mathcal{O}(\epsilon^2),\\
D_2=&\frac{3}{16}\int_0^\infty dc_{12} \int_0^\lambda d\tilde{b}\hspace{0.2em} 
		\tilde{b}c_{12}^7 (7+c_{12}^2) \sin^2\chi^{(0)} \exp\left(-\frac{1}{2}c_{12}^2\right)\nonumber\\
	&+\epsilon \left[
		\frac{3}{256}\int_0^\infty dc_{12} \int_0^{\tilde{b}_{\rm max}}d\tilde{b}\hspace{0.2em} 
		\tilde{b}(\nu^2-\tilde{b}^2)c_{12}^5 (-595-252c_{12}^2-18c_{12}^4+4c_{12}^6+c_{12}^8)
		\exp\left(-\frac{1}{2}c_{12}^2\right)\right.\nonumber\\
	&\hspace{2em}\left.+\frac{3}{4}\int_0^\infty dc_{12} \int_0^{\tilde{b}_{\rm max}}d\tilde{b}\hspace{0.2em} 
		\tilde{b}(\nu^2-\tilde{b}^2)c_{12}^7 (7-c_{12}^2) \sin^2\frac{\chi^{(0)}}{2}\exp\left(-\frac{1}{2}c_{12}^2\right)\right.\nonumber\\
	&\hspace{2em}\left.+\frac{3}{2}\int_0^\infty dc_{12} \int_0^{\tilde{b}_{\rm max}}d\tilde{b}\hspace{0.2em} 
		\tilde{b}(\nu^2-\tilde{b}^2)c_{12}^9 \sin^4\frac{\chi^{(0)}}{2}\exp\left(-\frac{1}{2}c_{12}^2\right)\right.\nonumber\\
	&\hspace{2em}\left.+\frac{3}{16}\int_0^\infty dc_{12} \int_0^\lambda d\tilde{b}\hspace{0.2em} 
		\tilde{b}c_{12}^7 (7+c_{12}^2) \chi^{(1)}\sin^2 2\chi^{(0)} \exp\left(-\frac{1}{2}c_{12}^2\right)\right]
	+\mathcal{O}(\epsilon^2),
\end{align}
\begin{align}
D_3=&\frac{3}{64}\int_0^\infty dc_{12} \int_0^\lambda d\tilde{b}\hspace{0.2em} 
	\tilde{b}c_{12}^7 (35-c_{12}^4) \sin^2\chi^{(0)} \exp\left(-\frac{1}{2}c_{12}^2\right)\nonumber\\
	&+\epsilon \left[
	\frac{1}{1024}\int_0^\infty dc_{12} \int_0^{\tilde{b}_{\rm max}}d\tilde{b}\hspace{0.2em} 
	\tilde{b}(\nu^2-\tilde{b}^2)c_{12}^5 (-5145-1785c_{12}^2+798c_{12}^4+22c_{12}^6+7c_{12}^8-c_{12}^{10})
	\exp\left(-\frac{1}{2}c_{12}^2\right)\right.\nonumber\\
	&\hspace{2em}\left.+\frac{3}{16}\int_0^\infty dc_{12} \int_0^{\tilde{b}_{\rm max}}d\tilde{b}\hspace{0.2em} 
	\tilde{b}(\nu^2-\tilde{b}^2)c_{12}^7 (35-14c_{12}^2+c_{12}^4) 
	\sin^2\frac{\chi^{(0)}}{2}\exp\left(-\frac{1}{2}c_{12}^2\right)\right.\nonumber\\
	&\hspace{2em}\left.+\frac{3}{8}\int_0^\infty dc_{12} \int_0^{\tilde{b}_{\rm max}}d\tilde{b}\hspace{0.2em} 
	\tilde{b}(\nu^2-\tilde{b}^2)c_{12}^9 (7-c_{12}^2)\sin^4\frac{\chi^{(0)}}{2}\exp\left(-\frac{1}{2}c_{12}^2\right)\right.\nonumber\\
	&\hspace{2em}\left. + \frac{3}{64}\int_0^\infty dc_{12} \int_0^\lambda d\tilde{b}\hspace{0.2em} 
	\tilde{b}c_{12}^7 (35-c_{12}^4) \chi^{(1)} \sin^2 2\chi^{(0)} \exp\left(-\frac{1}{2}c_{12}^2\right)\right]
	+\mathcal{O}(\epsilon^2).
\end{align}
Here, we only show the lowest order of each term.
Here, ${\cal M}_4$ and ${\cal M}_6$ are, respectively, related to ${\cal M}_2$,
the fourth moment $\left<c^4\right>$ and the sixth moment $\left<c^6\right>$ as
\begin{equation}
	\begin{cases}
	\displaystyle\frac{4}{3}{\cal M}_2 \left<c^4\right> = {\cal M}_4\\
	2 {\cal M}_2 \left<c^6\right> ={\cal M}_6
	\end{cases}.\label{eq:mu2mu4mu6}
\end{equation}
Substituting Eqs.\ (\ref{eq:mu246}) into Eq.\ (\ref{eq:mu2mu4mu6}) with $\left<c^4\right>=(15/4)(1+a_2)$ and $\left<c^6\right>=(105/8)(1+3a_2-a_3)$,
we obtain the simultaneous equations with respect to $a_2$ and $a_3$ as
\begin{equation}
	\begin{cases}
	\left(5S_1+5S_2-T_2\right)a_2 + \left(5S_3-T_3\right)a_3=T_1-5S_1\\
	\displaystyle\left(\frac{315}{4}S_1+\frac{105}{4}S_2-D_2\right)a_2 
		+ \left(-\frac{105}{4}S_1+\frac{105}{4}S_3-D_3\right)a_3=D_1-\frac{105}{4}S_1
	\end{cases}.
\end{equation}
These equations can be solved easily and the explicit forms of $a_2$ and $a_3$ up to $\epsilon$ order are
given by Eqs.\ (\ref{eq:a2a3})--(\ref{eq:D_a2a3}).
Thus, we explicitly write ${\cal M}_2$, ${\cal M}_4$, and ${\cal M}_6$ up to the first order of $\epsilon$ as
\begin{align}
{\cal M}_2 
=& \epsilon \sqrt{2\pi} \int_0^\infty dc_{12} \int_0^{\tilde{b}_{\rm max}} d\tilde{b}  
	\hspace{0.2em} \tilde{b}(\nu^2-\tilde{b}^2) c_{12}^5
	\exp\left(-\frac{1}{2}c_{12}^2\right)+\mathcal{O}(\epsilon^2),\label{eq:mu2_final}\\
{\cal M}_4 
=&\epsilon \left[\frac{\sqrt{2\pi}}{2}(1-e) \int_0^\infty dc_{12} \int_0^{\tilde{b}_{\rm max}} d\tilde{b}  
		\hspace{0.2em} \tilde{b} (\nu^2-\tilde{b}^2)c_{12}^5 \left(5+c_{12}^2\right)
		\exp\left(-\frac{1}{2}c_{12}^2\right)\right.\nonumber\\
	&\hspace{1em}\left.+a_2^{(1)} \frac{\sqrt{2\pi}}{4} \int_0^\infty dc_{12} \int_0^\lambda d\tilde{b}  
		\hspace{0.2em} \tilde{b} c_{12}^7 \sin^2 \chi^{(0)} \exp\left(-\frac{1}{2}c_{12}^2\right)\right.\nonumber\\
	&\hspace{1em}\left.+a_3^{(1)} \frac{\sqrt{2\pi}}{16} \int_0^\infty dc_{12} \int_0^\lambda d\tilde{b}  
		\hspace{0.2em} \tilde{b} c_{12}^7\left(7-c_{12}^2\right) \sin^2 \chi^{(0)} \exp\left(-\frac{1}{2}c_{12}^2\right)\right]
	+\mathcal{O}(\epsilon^2),\label{eq:mu4_final}\\
{\cal M}_6
=&\epsilon \left[\frac{3\sqrt{2\pi}}{16}(1-e) \int_0^\infty dc_{12} \int_0^{\tilde{b}_{\rm max}} d\tilde{b}  
		\hspace{0.2em} \tilde{b}(\nu^2-\tilde{b}^2) c_{12}^5 \left(35+14c_{12}^2+c_{12}^4\right)
		\exp\left(-\frac{1}{2}c_{12}^2\right)\right.\nonumber\\
	&\hspace{1em}\left.+a_2^{(1)} \frac{3\sqrt{2\pi}}{16} \int_0^\infty dc_{12} \int_0^\lambda d\tilde{b}  
		\hspace{0.2em} \tilde{b} c_{12}^7\left(7+c_{12}^2\right) \sin^2 \chi^{(0)} \exp\left(-\frac{1}{2}c_{12}^2\right)\right.\nonumber\\
	&\hspace{1em}\left.+a_3^{(1)} \frac{3\sqrt{2\pi}}{64} \int_0^\infty dc_{12} \int_0^\lambda d\tilde{b}  
		\hspace{0.2em} \tilde{b} c_{12}^7\left(35-c_{12}^4\right) \sin^2 \chi^{(0)} \exp\left(-\frac{1}{2}c_{12}^2\right)\right]
	+\mathcal{O}(\epsilon^2).
\end{align}

\section{Calculation of $\Omega_\eta^e$ and $\Omega_\kappa^e$}\label{sec:Omega}
In this appendix, we calculate the quantities $\Omega_\eta^e$ and $\Omega_\kappa^e$ 
in Eqs. (\ref{eq:Omega_eta_general}) and (\ref{eq:Omega_kappa_general}).
From the definition, $\tilde{D}_{ij}(\bm{c})=c_i c_j - c^2\delta_{ij}/3$, 
$\tilde{D}_{ij}(\bm{c}_2) \Delta \left[ \tilde{D}_{ij}(\bm{c}_1) + \tilde{D}_{ij}(\bm{c}_2)\right]$ 
is rewritten as
\begin{align}
&\tilde{D}_{ij}(\bm{c}_2) \Delta \left[ \tilde{D}_{ij}(\bm{c}_1) + \tilde{D}_{ij}(\bm{c}_2)\right]\nonumber\\
=&\left(c_{2i}c_{2j}-\frac{1}{3}\delta_{ij}c_2^2\right)
	\left[c_{1i}^\prime c_{1j}^\prime + c_{2i}^\prime c_{2j}^\prime -c_{1i}c_{1j}-c_{2i}c_{2j} -\frac{1}{3}\delta_{ij}\left(c_1^{\prime2}+c_2^{\prime2}-c_1^2-c_2^2\right) \right]\nonumber\\
=&c_{12}^2(\bm{C}\cdot \hat{\bm{k}})(\bm{c}_{12}\cdot \hat{\bm{k}})
	-\frac{1}{2}c_{12}^2 (\bm{c}_{12}\cdot \hat{\bm{k}})^2
	-2(\bm{C}\cdot \bm{c}_{12})(\bm{C}\cdot \hat{\bm{k}})(\bm{c}_{12}\cdot \hat{\bm{k}})
	+(\bm{C}\cdot \bm{c}_{12})(\bm{c}_{12}\cdot \hat{\bm{k}})^2\nonumber\\
	&+2(\bm{C}\cdot \hat{\bm{k}})^2(\bm{c}_{12}\cdot \hat{\bm{k}})^2
	-2(\bm{C}\cdot \hat{\bm{k}})(\bm{c}_{12}\cdot \hat{\bm{k}})^3
	+\frac{1}{2}(\bm{c}_{12}\cdot \hat{\bm{k}})^4\nonumber\\
&+\epsilon \hspace{0.2em}\Theta(\tilde{b}_{\rm max}-\tilde{b})\nu^2\frac{\cos^2\Theta}{\cos^2\theta}\left[
		\frac{1}{3}C^2(\bm{c}_{12}\cdot \hat{\bm{k}})^2
		-\frac{1}{2}c_{12}^2 (\bm{C}\cdot \hat{\bm{k}})(\bm{c}_{12}\cdot \hat{\bm{k}})
		+\frac{1}{3}c_{12}^2(\bm{c}_{12}\cdot \hat{\bm{k}})^2
		+(\bm{C}\cdot \bm{c}_{12})(\bm{C}\cdot \hat{\bm{k}})(\bm{c}_{12}\cdot \hat{\bm{k}})\right.\nonumber\\
	&\left.\hspace{12em}-\frac{5}{6}(\bm{C}\cdot \bm{c}_{12})(\bm{c}_{12}\cdot \hat{\bm{k}})^2
		-2(\bm{C}\cdot \hat{\bm{k}})^2(\bm{c}_{12}\cdot \hat{\bm{k}})^2
		+2(\bm{C}\cdot \hat{\bm{k}})(\bm{c}_{12}\cdot \hat{\bm{k}})^3
		-\frac{1}{2}(\bm{c}_{12}\cdot \hat{\bm{k}})^4
\right]\nonumber\\
&+\mathcal{O}(\epsilon^2).
\end{align}
Substituting this result into Eq.\ (\ref{eq:Omega_eta_general}), we obtain Eq.\ (\ref{eq:Omega_expand}).

For $\Omega_\kappa^e$, we rewrite $\tilde{\bm{S}}(\bm{c}_2) \cdot\Delta \left[ \tilde{\bm{S}}(\bm{c}_1) + \tilde{\bm{S}}(\bm{c}_2) \right]$ as
\begin{align}
&\tilde{\bm{S}}(\bm{c}_2) \cdot\Delta \left[ \tilde{\bm{S}}(\bm{c}_1) + \tilde{\bm{S}}(\bm{c}_2) \right]\nonumber\\
=& \left(c_2^2-\frac{5}{2}\right)
	\left[ \left(\bm{c}_1^\prime \cdot \bm{c}_2\right)c_1^{\prime2} 
	+ \left(\bm{c}_2^\prime \cdot \bm{c}_2\right)c_2^{\prime2}
	- \left(\bm{c}_1 \cdot \bm{c}_2\right)c_1^2 - \left(\bm{c}_2 \cdot \bm{c}_2\right)c_2^2 \right]\nonumber\\
=& C^2c_{12}^2 (\bm{C}\cdot \hat{\bm{k}}) (\bm{c}_{12}\cdot \hat{\bm{k}}) 
	-4C^2(\bm{C}\cdot \bm{c}_{12}) (\bm{C}\cdot \hat{\bm{k}}) (\bm{c}_{12}\cdot \hat{\bm{k}}) 
	+C^2 (\bm{C}\cdot \bm{c}_{12}) (\bm{c}_{12}\cdot \hat{\bm{k}}) ^2
	+4C^2 (\bm{C}\cdot \hat{\bm{k}})^2 (\bm{c}_{12}\cdot \hat{\bm{k}})^2\nonumber\\
&-2C^2 (\bm{C}\cdot \hat{\bm{k}}) (\bm{c}_{12}\cdot \hat{\bm{k}})^3
	+\frac{1}{4}c_{12}^4 (\bm{C}\cdot \hat{\bm{k}}) (\bm{c}_{12}\cdot \hat{\bm{k}}) 
	-2c_{12}^2(\bm{C}\cdot \bm{c}_{12}) (\bm{C}\cdot \hat{\bm{k}}) (\bm{c}_{12}\cdot \hat{\bm{k}}) 
	+\frac{1}{4}c_{12}^2(\bm{C}\cdot \bm{c}_{12}) (\bm{c}_{12}\cdot \hat{\bm{k}})^2\nonumber\\
&+c_{12}^2 (\bm{C}\cdot \hat{\bm{k}})^2 (\bm{c}_{12}\cdot \hat{\bm{k}})^2
	-\frac{1}{2}c_{12}^2 (\bm{C}\cdot \hat{\bm{k}}) (\bm{c}_{12}\cdot \hat{\bm{k}})^3
	-\frac{5}{2}c_{12}^2(\bm{C}\cdot \hat{\bm{k}}) (\bm{c}_{12}\cdot \hat{\bm{k}})
	+4(\bm{C}\cdot \bm{c}_{12})^2 (\bm{C}\cdot \hat{\bm{k}}) (\bm{c}_{12}\cdot \hat{\bm{k}}) \nonumber\\
&-(\bm{C}\cdot \bm{c}_{12})^2(\bm{c}_{12}\cdot \hat{\bm{k}})^2
	-4(\bm{C}\cdot \bm{c}_{12}) (\bm{C}\cdot \hat{\bm{k}})^2 (\bm{c}_{12}\cdot \hat{\bm{k}})^2
	+2(\bm{C}\cdot \bm{c}_{12}) (\bm{C}\cdot \hat{\bm{k}}) (\bm{c}_{12}\cdot \hat{\bm{k}})^3
	+10(\bm{C}\cdot \bm{c}_{12}) (\bm{C}\cdot \hat{\bm{k}}) (\bm{c}_{12}\cdot \hat{\bm{k}})\nonumber\\
&-\frac{5}{2}(\bm{C}\cdot \bm{c}_{12}) (\bm{c}_{12}\cdot \hat{\bm{k}})^2
	-10 (\bm{C}\cdot \hat{\bm{k}})^2 (\bm{c}_{12}\cdot \hat{\bm{k}})^2
	+5 (\bm{C}\cdot \hat{\bm{k}}) (\bm{c}_{12}\cdot \hat{\bm{k}})^3\nonumber\\
&+\epsilon \hspace{0.2em}\Theta(\tilde{b}_{\rm max}-\tilde{b})\nu^2\frac{\cos^2\Theta}{\cos^2\theta}\left[
	-C^4(\bm{c}_{12}\cdot \hat{\bm{k}})^2
	-\frac{1}{2}C^2c_{12}^2 (\bm{C}\cdot \hat{\bm{k}}) (\bm{c}_{12}\cdot \hat{\bm{k}})
	-\frac{1}{4}C^2c_{12}^2(\bm{c}_{12}\cdot \hat{\bm{k}})^2
	+2C^2(\bm{C}\cdot \bm{c}_{12}) (\bm{C}\cdot \hat{\bm{k}}) (\bm{c}_{12}\cdot \hat{\bm{k}})\right.\nonumber\\
&\hspace{5em}\left.+C^2(\bm{C}\cdot \bm{c}_{12}) (\bm{c}_{12}\cdot \hat{\bm{k}})^2
	-4C^2(\bm{C}\cdot \hat{\bm{k}})^2 (\bm{c}_{12}\cdot \hat{\bm{k}})^2
	+2C^2 (\bm{C}\cdot \hat{\bm{k}}) (\bm{c}_{12}\cdot \hat{\bm{k}})
	+\frac{5}{2}C^2(\bm{c}_{12}\cdot \hat{\bm{k}})^2\right.\nonumber\\
&\hspace{5em}\left.-\frac{1}{8}c_{12}^4 (\bm{C}\cdot \hat{\bm{k}}) (\bm{c}_{12}\cdot \hat{\bm{k}})
	+c_{12}^2(\bm{C}\cdot \bm{c}_{12}) (\bm{C}\cdot \hat{\bm{k}}) (\bm{c}_{12}\cdot \hat{\bm{k}})
	-c_{12}^2 (\bm{C}\cdot \hat{\bm{k}})^2 (\bm{c}_{12}\cdot \hat{\bm{k}})^2
	+\frac{1}{2}c_{12}^2 (\bm{C}\cdot \hat{\bm{k}}) (\bm{c}_{12}\cdot \hat{\bm{k}})^3\right.\nonumber\\
&\hspace{5em}\left.+\frac{5}{4}c_{12}^2 (\bm{C}\cdot \hat{\bm{k}}) (\bm{c}_{12}\cdot \hat{\bm{k}})
	-2(\bm{C}\cdot \bm{c}_{12})^2 (\bm{C}\cdot \hat{\bm{k}}) (\bm{c}_{12}\cdot \hat{\bm{k}})
	+4(\bm{C}\cdot \bm{c}_{12}) (\bm{C}\cdot \hat{\bm{k}})^2 (\bm{c}_{12}\cdot \hat{\bm{k}})^2\right.\nonumber\\
&\hspace{5em}\left.-2(\bm{C}\cdot \bm{c}_{12}) (\bm{C}\cdot \hat{\bm{k}}) (\bm{c}_{12}\cdot \hat{\bm{k}})^3
	-5(\bm{C}\cdot \bm{c}_{12}) (\bm{C}\cdot \hat{\bm{k}}) (\bm{c}_{12}\cdot \hat{\bm{k}})
	+10 (\bm{C}\cdot \hat{\bm{k}})^2 (\bm{c}_{12}\cdot \hat{\bm{k}})^2
	-5(\bm{C}\cdot \hat{\bm{k}}) (\bm{c}_{12}\cdot \hat{\bm{k}})^3
\right]\nonumber\\
	&+\mathcal{O}(\epsilon^2).
\end{align}
Substituting this into Eq.\ (\ref{eq:Omega_kappa_general}), we obtain Eq.\ (\ref{eq:Omega_expand}) after the long and tedious calculation.

\end{widetext}
\section{High and low temperature expansions}\label{sec:highTexpansion}
We can evaluate the explicit forms of the transport coefficients in terms of high temperature expansion.
We can also evaluate the dissipation rate ${\cal M}_2$ as a low temperature expansion, though it diverges in the low temperature limit.

First, we discuss the high temperature expansion.
From Eq.\ (\ref{eq:def_nu}), we expand $\nu$ as
\begin{align}
	\nu=\sqrt{1+\frac{2\varepsilon}{T c_{12}^2}}
	=1+\frac{\varepsilon}{T} \frac{1}{c_{12}^2} + \mathcal{O}\left(\left(\frac{\varepsilon}{T}\right)^2\right),\label{eq:nu_expansion}
\end{align}
for $T/\varepsilon \gg 1$.
Substituting Eq.\ (\ref{eq:nu_expansion}) into Eq.\ (\ref{eq:mu2_expand}), we expand ${\cal M}_2$ in terms of the small parameter $\varepsilon/T$ as
\begin{equation}
	{\cal M}_2^{(0)}= 0,\quad
	{\cal M}_2^{(1)}= {\cal M}_2^{(1,0)}+\frac{\varepsilon}{T} {\cal M}_2^{(1,1)} + \mathcal{O}\left(\left(\frac{\varepsilon}{T}\right)^2\right)
	\label{eq:mu_2_expansion}
\end{equation}
with
\begin{equation}
	{\cal M}_2^{(1,0)}= 2\sqrt{2\pi},\quad
	{\cal M}_2^{(1,1)}= 2\sqrt{2\pi}.
\end{equation}
Similarly, $\Omega_\eta^e$ and $\Omega_\kappa^e$ are, respectively, expanded as
\begin{align}
	\Omega_\eta^{e(0)}&= \Omega_\eta^{e(0,0)} + \frac{\varepsilon}{T}\Omega_\eta^{e(0,1)} 
						+ \mathcal{O}\left(\left(\frac{\varepsilon}{T}\right)^2\right),\label{eq:Omega_eta_e0_expansion}\\
	\Omega_\eta^{e(1)}&= \Omega_\eta^{e(1,0)} + \mathcal{O}\left(\frac{\varepsilon}{T}\right),\label{eq:Omega_eta_e1_expansion}\\
	\Omega_\kappa^{e(0)}&= \Omega_\kappa^{e(0,0)} + \frac{\varepsilon}{T}\Omega_\kappa^{e(0,1)} 
						+ \mathcal{O}\left(\left(\frac{\varepsilon}{T}\right)^2\right),\\
	\Omega_\kappa^{e(1)}&= \Omega_\kappa^{e(1,0)} + \mathcal{O}\left(\frac{\varepsilon}{T}\right)\label{eq:Omega_kappa_e1_expansion}
\end{align}
with
\begin{align}
	\Omega_\eta^{e(0,0)} &= -4\sqrt{2\pi},\quad
	\Omega_\eta^{e(1,0)} = -\frac{11\sqrt{2\pi}}{320},\\
	\Omega_\eta^{e(0,1)} &= \frac{\sqrt{2\pi}}{24}(\lambda-1)
		\left\{2(15\lambda^4+15\lambda^3+2\lambda^2+2\lambda+2)\right.\nonumber\\
		&\hspace{3em}\left.+3\lambda^2(\lambda+1)(5\lambda^2-1)\log \frac{\lambda-1}{\lambda+1}\right\},\\
	\Omega_\kappa^{e(0,0)} &= -4\sqrt{2\pi},\quad
	\Omega_\kappa^{e(1,0)} = -\frac{1989\sqrt{2\pi}}{320},\\
	\Omega_\kappa^{e(0,1)} &= \frac{\sqrt{2\pi}}{24}(\lambda-1)
		\left\{2(15\lambda^4+15\lambda^3+2\lambda^2+2\lambda+2)\right.\nonumber\\
		&\hspace{3em}\left.+3\lambda^2(\lambda+1)(5\lambda^2-1)\log \frac{\lambda-1}{\lambda+1}\right\}.
\end{align}

Next, let us calculate the expansions of the transport coefficients.
Substituting Eqs.\ (\ref{eq:mu_2_expansion})--(\ref{eq:Omega_eta_e1_expansion}) into Eqs.\ (\ref{eq:eta_0}) and (\ref{eq:eta_1}), 
we expand $\eta$ as
\begin{align}
	\eta^{(0)}&= \eta^{(0,0)} + \frac{\varepsilon}{T}\eta^{(0,1)} +\mathcal{O}\left(\left(\frac{\varepsilon}{T}\right)^2\right),\\
	\eta^{(1)}&= \eta^{(1,0)} + \mathcal{O}\left(\frac{\varepsilon}{T}\right)
\end{align}
with
\begin{align}
	\eta^{(0,0)}&=\frac{5}{16d^2}\sqrt{\frac{mT}{\pi}},\quad
	\eta^{(1,0)}= \frac{1567}{3840}\eta^{(0,0)},\\
	\eta^{(0,1)}&= \eta^{(0,0)} \frac{\lambda-1}{96}\left\{2(15\lambda^4+15\lambda^3+2\lambda^2+2\lambda+2)\right.\nonumber\\
				&\hspace{3em}\left.+3\lambda^2(\lambda+1)(5\lambda^2-1)\log \frac{\lambda-1}{\lambda+1}\right\}.
\end{align}
Similarly, $\kappa$ and $\mu$ are, respectively, expanded as
\begin{align}
	\kappa^{(0)}&= \kappa^{(0,0)} + \frac{\varepsilon}{T}\kappa^{(0,1)} +\mathcal{O}\left(\left(\frac{\varepsilon}{T}\right)^2\right),\\
	\kappa^{(1)}&= \kappa^{(1,0)} + \mathcal{O}\left(\frac{\varepsilon}{T}\right),\\
	\mu^{(0)}&=0,\quad \mu^{(1)}= \mu^{(1,0)} + \mathcal{O}\left(\frac{\varepsilon}{T}\right)
\end{align}
with
\begin{align}
	\kappa^{(0,0)}&=\frac{75}{64d^2}\sqrt{\frac{T}{\pi m}},\quad
	\kappa^{(1,0)}=\frac{539}{1280}\kappa^{(0,0)},\\
	\kappa^{(0,1)}&=\kappa^{(0,0)}\frac{\lambda-1}{96}\left\{2(15\lambda^4+15\lambda^3+2\lambda^2+2\lambda+2)\right.\nonumber\\
				&\hspace{3em}\left.+3\lambda^2(\lambda+1)(5\lambda^2-1)\log \frac{\lambda-1}{\lambda+1}\right\},\\
	\mu^{(1,0)}&=\frac{1185}{1024nd^2}\sqrt{\frac{T^3}{\pi m}}.
\end{align}

Let us also calculate the low temperature expansion of ${\cal M}_2$.
From Eq.\ (\ref{eq:def_nu}), we expand $\nu$ as
\begin{equation}
	\nu 	=\frac{\sqrt{2}}{c_{12}}\sqrt{\frac{\varepsilon}{T}} + \frac{\sqrt{2}c_{12}}{4}\frac{T}{\varepsilon}
		+\mathcal{O}\left(\left(\frac{T}{\varepsilon}\right)^3\right).\label{eq:nu_low_T_expansion}
\end{equation}
Substituting Eq.\ (\ref{eq:nu_low_T_expansion}) into Eq.\ (\ref{eq:mu2_expand}), we can expand ${\cal M}_2$ in terms of the small parameter $T/\varepsilon$ as
\begin{equation}
	{\cal M}_2^{(0)}= 0,\quad
	{\cal M}_2^{(1)}= \frac{\varepsilon}{T}{\cal M}_{2,0}^{(1,-1)}+ {\cal M}_{2,0}^{(1,0)} + \mathcal{O}\left(\sqrt{\frac{T}{\varepsilon}}\right)
	\label{eq:mu_2_low_T_expansion}
\end{equation}
with
\begin{equation}
	{\cal M}_{2,0}^{(1,-1)}=2\sqrt{2\pi}\lambda^2,\quad
	{\cal M}_{2,0}^{(1,0)}=2\sqrt{2\pi}.
\end{equation}


\section{DSMC algorithm}\label{sec:DSMC_algorithm}
In this appendix, we briefly summarize the DSMC procedure \cite{Bird, Garcia,Nanbu1980,Nanbu1983, Poschel},
which is a numerical technique to obtain the solution of the Boltzmann equation at $t+\Delta t$ from that at $t$.
For small $\Delta t$, the velocity distribution function at $t+\Delta t$ is given by
\begin{align}
	f(\bm{v},t+\Delta t)= f(\bm{v},t) + \frac{\partial f(\bm{v},t)}{\partial t}\Delta t.
\end{align}
Substituting the Boltzmann equation (\ref{eq:BE}) into this, we obtain
\begin{align}
	f(\bm{v},t+\Delta t)&= \left(1-\Delta t D + \Delta t J\right)f(\bm{v},t)\nonumber\\
	&= \left(1+\Delta t J\right)\left(1-\Delta t D\right)f(\bm{v},t) + \mathcal{O}\left(\Delta t^2\right),\label{eq:BE_DSMC}
\end{align}
where we have introduced $Df = \bm{v}\cdot \bm\nabla f$ and $Jf=I(f,f)$ for simplicity.
Equation (\ref{eq:BE_DSMC}) shows that the time evolution of the velocity distribution function can be separated into two parts: advective process and collision process.
According to this separation, DSMC iteration is as follows:
(i) We determine the time step $\Delta t$ smaller than $L/v_{\rm max}$, where $L$ is the system size and $v_{\rm max}$ is the maximum speed among the particles, which is evaluated as $v_{\rm max}=5v_{\rm T}$ with the thermal velocity $v_{\rm T}$. In this paper, we adopt $\Delta t=0.2L/v_{\rm max}$.
(ii) We move the particles during $\Delta t$ without any collisions. 
This corresponds to update the distribution function $f^*(\bm{v},t)=\left(1-\Delta t D\right)f(\bm{v},t)$.
(iii) We modify the velocities of the particles due to collisions.
We randomly determine the collisions without taking into account the actual positions of the particles.
The square of the collision parameter, $b^2$, is chosen in the range $0<b^2<\lambda^2 d^2$ at random.
A pair of colliding particles change the velocities according to rule in Eqs.\ (\ref{eq:vel_change}) and (\ref{eq:vel_change1}) for a hard core collisions and 
Eqs.\ (\ref{eq:vel_change}) and (\ref{eq:vel_change2}) for a grazing collision
Here, the number of collisions $N_c$ is evaluated as $\pi (\lambda d)^2N^2v_{\rm max}\Delta t$, which is proportional to the total cross section, the maximum speed, and the time step $\Delta t$.
This process corresponds to obtain $f(\bm{v},t+\Delta t)= \left(1+\Delta t J\right)f^*(\bm{v},t)$.
(iv) We update the time $t+\Delta t$.


\section{Estimation of the trapping temperature}\label{sec:T_escape}
As stated in Sec.\ \ref{sec:deflection}, we ignore the trapping process by the potential well throughout the paper.
In this Appendix, we briefly discuss the critical condition which validates this approximation using a simple one-dimensional model.
Let us consider a process in which two particles approach from far away relative speed $v$ in the frame that the target is stationary.
When the particle enters the potential region ($d<r<\lambda d$), the velocity becomes 
\begin{equation}
v_{\rm in}=\sqrt{v^2+\frac{4\varepsilon}{m}}\label{eq:v_in}
\end{equation}
from the energy conservation $(1/2)m_{\rm r}v^2=(1/2)m_{\rm r}v_{\rm in}^2 - \varepsilon$ with the reduced mass $m_{\rm r}=m/2$.
After the inelastic scattering on the hard core ($r=d$), the velocity changes from $v_{\rm in}$ to $-ev_{\rm in}$.
When the particle is trapped by the potential, the energy is negative, that is, $(1/2)m_{\rm r}e^2 v_{\rm in}^2 - \varepsilon <0$.
Using Eq.\ (\ref{eq:v_in}), the trapping condition is given by
\begin{equation}
	v<v_{\rm trap}=\sqrt{\frac{8\varepsilon}{m}(1-e)} + \mathcal{O}(1-e).
\end{equation}
The corresponding granular temperature is given by
\begin{equation}
	T<T_{\rm trap}=\frac{1}{2}m v_{\rm trap}^2=4\varepsilon (1-e)+\mathcal{O}\left((1-e)^2\right).
\end{equation}
The critical trapping temperature is given by $T_{\rm trap}\simeq 0.04\varepsilon$ for $e=0.99$, which is consistent with the lower bound of our theory as shown in Figs.\ \ref{fig:eta_kin} and \ref{fig:kappa_prime}.



\end{document}